\documentclass[aps, pra, amsfonts, amssymb, amsmath, showkeys, twocolumn, nofootinbib, 10pt]{revtex4-1}
\usepackage{lmodern}

\usepackage[english]{babel}
\usepackage[T1]{fontenc}
\usepackage[utf8]{inputenc}
\usepackage[bottom]{footmisc}
\usepackage{physics}
\usepackage{esint}
\usepackage{color}
\usepackage{graphicx}
\usepackage{geometry}
\usepackage{float}
\usepackage[unicode=true,pdfusetitle,
bookmarks=true,bookmarksnumbered=false,bookmarksopen=false,
 breaklinks=false,pdfborder={0 0 1},backref=false,colorlinks=true]
 {hyperref}
\hypersetup{citecolor=blue,linkcolor=blue,urlcolor=blue}
\usepackage{float}
\usepackage{multirow}

\begin{document}

\title{\texorpdfstring{The effective phase space and $e$-folding of the Starobinsky and extended Starobinsky model of inflation}{The effective phase space and e-folding of the Starobinsky and extended Starobinsky model of inflation}}
\author{Theppawan Rukpakawong}
    \email{theppawan.k@gmail.com}
    \affiliation{High Energy Physics Theory Group, Department of Physics, Faculty of Science, Chulalongkorn University, Bangkok 10330, Thailand}
\author{Piyabut Burikham}
    \email{piyabut@gmail.com}
    \affiliation{High Energy Physics Theory Group, Department of Physics, Faculty of Science, Chulalongkorn University, Bangkok 10330, Thailand}
\date{\today}

\begin{abstract}
For zero spatial curvature, cosmological phase space of Starobinsky and extended Starobinsky inflationary model show three apparent attractors; the fixed angle attractor in the large field limit, the final attractor representing reheating phase in the small field region, and the apparent attractor corresponding to the slow-roll condition connecting between the large-field and small-field region. To consider the total $e$-folding likelihood of the model, Remmen-Carroll conserved measure is constructed and normalized. Using the measure, the total e-folding number $N$ and its expectation value $\left\langle N \right\rangle$ are calculated. Our results show that most classical slow-roll trajectories which intersect the Planck surface have $N<60$, and $\phi_{\rm UV}>5.5 M^{*}_{\rm Pl}$ is required for $N>60$. It is found that for $\phi_{\rm UV}\in [5.22,5.50]M^{*}_{\rm Pl}$ which satisfies the constraint on the spectral index, $n_s = 0.9658 \pm 0.0040\,\,\,(68\%\,\,{\rm CL})$, the expectation value $\langle N \rangle \simeq 3.5 - 4$ for trajectories intersecting the Planck surface in the Starobinsky model. For extended Starobinsky model with additional $R^3$ term parametrized by a coupling parameter $\alpha$, the expectation value when inflation starts from the top of the potential shifts to $\left\langle N \right\rangle = 4.025,4.336$ for $\alpha = 10^{-4},6.5\times 10^{-5}$ respectively. In the Starobinsky model even at very large inflaton cutoff $\phi_{\rm UV}$ where the field value is super-Planckian, the energy density from the (saturating) inflaton potential and the Hubble parameter are still sub-Planckian and therefore the inflation occurs within the semi-classical regime. 
\end{abstract}

\maketitle

\section{Introduction} \label{sec:introduction}

Inflation provides a natural mechanism~\cite{Linde:1981mu,STAROBINSKY1982175,Cosmo_perturbation_Brandenberger_1983,Maldacena_2003} to explain the correlation we observed in the Cosmic Microwave Background Radiation~(CMB) and matter distribution over the entire sky, the observable universe appears to be homogeneous in the cosmic scale and in the early stage. Inflation was considered to also explain the spatial flatness of the universe~\cite{Guth:1980zm,Linde:1981mu}. However, study of the cosmological phase space in the Friedmann-Lemaitre-Robertson-Walker~(FLRW) universe reveals that the flat FLRW universe is actually common with the GHS~(Gibbons-Hawking-Stewart) phase space measure~\cite{Gibbons:1986xk} being divergent at zero spatial curvature~\cite{carroll2010unitaryevolutioncosmologicalfinetuning}. To avoid the infinite measure at zero curvature in the full GHS phase space, Remmen and Carroll~\cite{Remmen_2013,Remmen_2014} proposed an effective invariant phase space measure when the spatial curvature is fixed to zero. With the invariant measure, the probability for inflation with particular total $e$-folds and the expectation value of the total $e$-folds for a specific model of inflation can be evaluated. 

Starobinsky model~\cite{STAROBINSKY198099}~(see Ref.~\cite{Bianchi:2024qyp,Toyama:2024ugg} for recent review in Einstein and Jordan frame) was proposed as an effective quantum gravity action containing one-loop quantum gravity correction. It can be categorized as a class of $f(R)$ gravity theory that contains quadratic term of the Ricci scalar~\cite{Nojiri:2017ncd,Odintsov:2017hbk,Sebastiani:2013eqa,Myrzakulov:2014hca,Canko:2019mud}. The $f(R)$ gravity can be transformed further into a single-scalar-field inflaton theory. In the very early stage of the universe when the energy of the field is close to the Planck scale, quantum gravity fluctuations are expected to be significant so it is natural to consider quantum gravity effects in the inflationary period. Remarkably, Starobinsky model demonstrates that quantum gravity effects generate inflation.

Since the Starobinsky model considers only the leading correction to the Einstein-Hilbert action, it is natural to make extension to include higher-order quantum gravity contributions and other effects. There is an extensive literature on the extension models, which includes; the modular invariant Starobinsky inflation~\cite{Casas:2024jbw}, the modified singularity-free Starobinsky model with $R^{4}$-term~\cite{Saburov:2024und}, the machine-learning improved Starobinsky model~\cite{Kamerkar:2022dfu}, the scalar-$R^{2}$ coupling model~\cite{Chaichian:2022apa}, the cubic geometric inflation~\cite{ARCINIEGA2020135272}, the supersymmetric extension models~(see Ref.~\cite{Ketov:2012yz,Linde:2014nna} and references therein), the polynomial $f(R)$ inflation model~\cite{Huang_2014}, $R^p$ inflation~\cite{Motohashi_2015}, the unimodular mimetic $f(R)$ inflation~\cite{Odintsov_Oikonomou_2016}, and the extended Starobinsky model containing the effects of $R^3$ and $R\Box R$~\cite{Berkin_Maeda_1990}. 

Among the $f(R)$ gravity extension of the Starobinsky model, an addition of $R^3$ term with small coupling has proven to be phenomenologically viable~\cite{Cheong:2020rao, Modak:2022gol, Ivanov_2022, Burikham_2024}. In recent years, a large number of observational data becomes available, Planck CMB 2018 (TTTEEE+lowE+lensing)~\cite{Planck:2018jri,Planck:2018vyg}, BK18 (Bicep Keck 2018)~\cite{BICEP:2021xfz}, large-scale structure data, such as BAO (Baryon Acoustic Oscillation)~\cite{BOSS:2016wmc} and DES (Dark Energy Servey)~\cite{DES:2017myr,DES:2021bvc}. Observational constraints from especially Planck CMB, BAO and BK18 while ruling out a number of inflationary models, are consistent with the Starobinsky model. In Ref.~\cite{Burikham_2024}, it is shown that upper bound on e-folds on horizon exit can be estimated from reheating constraints which gives $N_e < 55-59$ for $k_*=0.05,0.002$ Mpc$^{-1}$ respectively, but the actual total e-folds are not constrained. However, in the extended Starobinsky model with presence of a cubic term $R^3$~\cite{Cheong:2020rao,Ivanov_2022}, all observational constraints cannot put a bound on the upper limit of $N_e$ without imposing certain reheating conditions. It is thus interesting to consider the likelihood of various total numbers of e-folding in the Starobinsky and extended Starobinsky model. In this work, we adopt the Remmen-Carroll conserved measure in subspace of cosmological dynamical phase space at zero spatial curvature to analyze the likelihood of various total e-folds in the two models. 

This work is organized as the following. In Sect.~\ref{sec:develop}, we review the effective phase space proposed by Ref.~\cite{Remmen_2013,Remmen_2014} and summarize the framework of slow-roll inflation. In Sect.~\ref{working1}, we analyze the effective phase space and construct the conserved measure in the Starobinsky model. We then compute the $e$-folds under the slow-roll approximation, evaluate the probability distribution on the space of trajectories, and calculate both the expectation value of $e$-folding number across all possible cosmological conditions and the probability that inflation lasts longer than 50 or 60 $e$-folds. These analyses are also applied to the extended Starobinsky model in Sect.~\ref{sec:working2}. Finally, our findings are summarized in Sect.~\ref{Conclusion}.

\section{the effective phase space of Inflation} \label{sec:develop}

In this section, we briefly review the simplest model of inflation and the associated cosmological phase space. We assume the universe is spatially homogeneous and isotropic. The accelerated expansion of the universe is driven by a single canonical scalar field. The action of scalar field coupled to general relativity on an expanding FLRW universe in the Einstein frame is~(see e.g. \cite{Burikham_2024} and references therein)
\begin{equation}\label{Ein_action}
    S = \frac{c^3}{2\kappa^2}\int d^4x\sqrt{-g}\left[R-g^{\mu\nu}\partial_\mu s \partial_\nu s - V(s)\right].
\end{equation}
Henceforth, we will use the natural unit with $c=\hbar=1$. We also define $1/\kappa \equiv M^{*}_{\mathrm{Pl}}=2.4 \times 10^{18}$ GeV and the dimensionless scalar field $s \equiv \phi/M^{*}_{\mathrm{Pl}}$. The action now can be written as
\begin{equation}
    S = \int d^4x\sqrt{-g}\left[\frac{{M^{*}_{\mathrm{Pl}}}^2}{2}R - \frac{1}{2}g^{\mu\nu}\partial_\mu \phi \partial_\nu \phi - V(\phi)\right],
\end{equation}
where $\phi$ is an inflaton field. In Hamiltonian analysis, we adopt the Arnowitt-Deser-Misner (ADM) formalism~\cite{Arnowitt_Deser_Misner_2008} to express the spacetime metric in a 3+1 decomposition. This involves slicing space into constant-time sections and using a lapse function to establish time-evolution~\cite{Pebralia2019InvestigatingSI,corichi2023introductionadmformalism}. The FLRW metric with the lapse function can be written as~\cite{Remmen_2014}
\begin{equation}
    ds^2 = -N^2(t)dt^2 + a^2(t)\left(\frac{dr^2}{1-k r^2} +r^2d\Omega^2 \right),
\end{equation}
where $N(t)$ is the lapse function and the curvature parameter $k \in \mathbb{R}$ is fixed for a given FLRW universe. Taking $\phi(t)$ as a function of time, the Hamiltonian is given by
\begin{equation}\label{Hamiltonian}
    \mathcal{H} = N\left[-\frac{p_a^2}{12a{M^{*}_{\rm Pl}}^2}+\frac{p_\phi^2}{2a^3}+a^3V(\phi)-3a k {M^{*}_{\rm Pl}}^2\right],
\end{equation}
where $p_a = -6N^{-1}{M^{*}_{\rm Pl}}^2a\dot{a}$, $p_\phi = N^{-1}a^3\dot{\phi}$, and $p_N = 0$ since $N$ is a Lagrange multiplier~\cite{Remmen_2013} to only impose the constraint. After evaluating Hamilton's equation with respect to $\phi$, we obtain the equation of motion of the scalar:
\begin{equation}\label{eqn:motion}
    \ddot{\phi} + 3H\dot{\phi} +V'(\phi) = 0,
\end{equation}
where the Hubble parameter $H \equiv \dot{a}/a$. The vector field in $\phi-\dot{\phi}$ space is 
\begin{equation}\label{eqn:traj}
    \mathbf{v} = \left( \dot{\phi}, -3H\dot{\phi}-V'(\phi)\right),
\end{equation}
where $V'(\phi) = {\rm d} V/{\rm d}\phi$. Furthermore, varying the action with respect to $N$ gives the Hamiltonian constraint, which is equivalent to the Friedmann equation, describing how the inflaton field affects the expansion of our universe:
\begin{equation}\label{eqn:Friedmann}
    H^2 = \frac{1}{3{M^{*}_\mathrm{Pl}}^2}\left[\frac{1}{2}\dot{\phi}^2 + V(\phi)\right]-\frac{k}{a^2}.
\end{equation}
Taking a time derivative on the Friedmann equation~\eqref{eqn:Friedmann} and then combining to scalar equation of motion~\eqref{eqn:motion}, to obtain
\begin{equation}\label{eqn:hubble_rate}
    \dot{H} = -\frac{1}{2}\frac{\dot{\phi}^2}{{M_{\rm Pl}^{*}}^2}.
\end{equation}

However, an analysis of the phase space on cosmic inflation using GHS measure~\cite{Gibbons:1986xk} reveals that the measure diverges with factor $|\Omega_k|^{-5/2}$ as the spatial curvature, $k$, approaches zero~\cite{carroll2010unitaryevolutioncosmologicalfinetuning}. Nearly all trajectories correspond to a flat universe. Consequently, analysis of the cosmological phase space focuses on finding a measure in the space of a flat universe, when $k = 0$; it leads to the scale factor, $a$, vanishing from the equation of motion~\eqref{eqn:motion} and~\eqref{eqn:Friedmann}. The equation of motion for inflation is then completely expressed by two variables: $\phi$ and $\dot{\phi}$.

In cosmology, it is often preferred to describe certain conditions or behaviors as \textit{natural}. In the context of Hamiltonian analysis, this perspective is reflected in the concept of a cosmological \textit{attractor} within phase space. Attractor behavior refers to scenarios where a group of points in phase space converge into a specific region and remain there indefinitely. This means that the evolution of trajectories reaches a certain behavior without requiring finely tuned initial conditions. However, for an attractor to exist, phase space volumes must contract in some regions. This contradicts the Liouville's theorem on the Hamiltonian system~\eqref{Hamiltonian} which states that the volume of a region in phase space remains constant over time, as discussed in~\cite{Gibbons_2008, Corichi_2011, Ashtekar_2011}.

To reconcile these two common characteristics in the cosmological phase space: attractor and Liouville's theorem, Remmen and Carroll~\cite{Remmen_2013} proposed the formalism for defining the conserved measure on the effective phase space $\Phi$ with non-canonical coordinates $\phi$ and $\dot{\phi}$ for universes with zero spatial curvature. This effective phase space can illustrate the appearance of certain attractor-like behavior by identifying it with the divergence of the measure and/or the negativity of Lyapunov exponents along certain coordinates. Furthermore, they found that the Liouville's measure ${\rm d}\pi_\phi \wedge {\rm d}\phi$ on effective phase space is just equivalent to constructing a two-form $f(\phi, \dot{\phi})\,{\rm d}\phi \wedge {\rm d}\dot{\phi}$ conserved under Hamiltonian flows~\footnote{defining the conjugate momentum on $\Phi$ as $\pi_\phi \equiv \partial \mathcal{L}_\Phi/\partial \dot{\phi}$, where $\mathcal{L}_\Phi$ is a Lagrangian description of the trajectories on the effective phase space $\Phi$. Ref.~\cite{Remmen_2013} also shows the condition for the existence of $\mathcal{L}_\Phi$.}.

\subsection{The probability distribution of different FLRW trajectories for universes}
Following Ref.~\cite{Remmen_2014}, the formalism of effective phase space have been applied in a wide range of potentials to construct a conserved measure and probability distribution of trajectories on the Planck surface ($H = M^{*}_{\rm Pl}$).  The $e$-folds average and the probability for particular total $e$-folds in a specific model of inflation can then be evaluated. 

The effective phase space for a given potential $V(\phi)$ is characterized by coordinates $(x,y)$, where $x,y$ are reparametrizing functions of $\phi,\dot{\phi}$ variables, respectively; and they can be expressed as
\begin{align}
    x(\phi) &\equiv \lambda \frac{\sqrt{V(\phi)}}{\sqrt{3}M^{*}_{\rm Pl}} \\
    y(\dot{\phi}) &\equiv \lambda \frac{\dot{\phi}}{\sqrt{6}M^{*}_{\rm Pl}},
\end{align}
where $\lambda$ is defined as an auxiliary parameter with mass dimension $-1$. By the definition of $x,y$ in Cartesian coordinates, we also can define polar coordinates $(z,\theta)$ in the effective phase space as
\begin{align}
    z &\equiv \sqrt{x^2+y^2} = \lambda H, \\
    \tan\theta &\equiv \frac{y}{x} = \frac{1}{\sqrt{2}}\frac{\dot{\phi}}{\sqrt{V(\phi)}}.
\end{align}
Hence, Hubble parameter $H$ is proportional to the radial component in polar coordinates. This setup clearly distinguishes the early universe from the late universe. 

Considering Liouville's theorem in this effective phase space, the conserved measure under evolution is a two-form
\begin{equation}
    \boldsymbol{\omega} = \omega(x,y)\,{\rm d}x \wedge {\rm d}y,
\end{equation}
which satisfies the conservation condition in context of the Lie derivative, $\pounds_{\mathbf{v}}\boldsymbol{\omega}=0$, where the vector field $\mathbf{v}$ is defined as components $(\dot{x}, \dot{y})$. Using the property of Lie derivation, we can equivalently write in component form as
\begin{equation}\label{conservation}
    \partial_\mu\left(\omega v^\mu\right)=0.
\end{equation}
The conservation~\eqref{conservation} will be used to evaluate the function of $\omega$ in term of coordinates in the effective phase space. Moreover, Liouville's theorem ensures the conservation of the volume element $\omega(x,y){\rm dx}{\rm d}y$ along a bundle of trajectories evolving over time, thereby providing the relation to a well-defined probability measure defined over the same bundle of trajectories. Hence, the probability distribution on the space of trajectories, parametrized by the coordinate $\theta$ which the trajectory intersects the surface of constant $H$, can be expressed as~\cite{Remmen_2014}
\begin{equation}\label{pdf}
    P(\theta)|_{H} = \frac{\omega(H,\theta)|\dot{H}|}{\int \omega(H,\theta')|\dot{H}|\mathrm{d}\theta'},
\end{equation}
where $\int \omega(H,\theta')|\dot{H}|\mathrm{d}\theta'$ is a normalization factor. This is an essential formula for evaluating the average number of $e$-folds, the fraction of the universe having more than 50 or 60 $e$-folds, and the probability distribution of inflation achieving a specific total number of $e$-folds.  

\subsection{The slow-roll inflation}\label{Sectslowroll}
During inflation, the comoving Hubble radius $(aH)^{-1}$ shrinking implies accelerated phase $\ddot{a}>0$. This phase occurs if Hubble slow-roll parameter $\epsilon < 1$. In order to have a long enough period of inflation to solve the horizon problem, we need to introduce the second Hubble slow-roll parameter $\eta$ to measure, and then inflation persists if $|\eta| < 1$. The slow-roll parameter $\epsilon$, $\eta$ are defined as follows:
\begin{equation}\label{hubble_slow-roll_parameter}
    \epsilon \equiv -\frac{\dot{H}}{H^2}, \quad \text{and} \quad 
    \eta \equiv \frac{\dot{\epsilon}}{H\epsilon}.
\end{equation}
Together with the expression for $\dot{H}$ given in Eqn.~\eqref{eqn:hubble_rate}, we can write the Friedmann equation and the scalar equation of motion in terms of $\eta$, $\epsilon$, and $V(\phi)$:
\begin{align}
    \label{eqn:Fried_w_parameter}
    H^2 &= \frac{1}{(3-\epsilon)M^{*2}_{\mathrm{Pl}}}V(\phi),  \quad \mathrm{and} 
    \\
    \label{eqn:EoM_w_parameter}
    3H\dot{\phi} &= -V'(\phi)\left(1+ \frac{\eta+2\epsilon}{6}\right)^{-1}.
\end{align}
We are looking at quasi-de Sitter expansion, where Hubble parameter $H$ is slowly declining with a sufficiently long time, during inflation. It is characterized by $\{\epsilon, |\eta| \ll 1\}$. We apply these conditions to simplify the scalar equation of motion and Friedmann equation. The slow-roll inflation is characterized by two physical conditions 
\begin{equation}\label{slow_roll_condition}
    \dot{\phi}^2 \ll |V(\phi)| \quad \mathrm{and} \quad |\ddot{\phi}| \ll |H\dot{\phi}|,|V'(\phi)|.
\end{equation}
At the leading-order slow-roll approximation, the Friedmann equation and the scalar equation of motion become
\begin{equation}\label{slow-roll behavior}
    H^2 \simeq \frac{1}{3M^{*2}_{\mathrm{Pl}}} V(\phi) \quad \mathrm{and} \quad 3H\dot{\phi} \simeq -V'(\phi),
\end{equation}
respectively. That is, the expansion of the universe is completely dependent on the potential energy, and the speed of the inflaton is determined by the field-gradient of the potential. For a given potential $V(\phi)$ in the single-inflaton model, the slow-roll approximation typically gives to the leading order,
\begin{equation}\label{slow-roll par}
    \epsilon=\epsilon_V \equiv \frac{M^{*2}_{\mathrm{Pl}}}{2}\left[\frac{V'(\phi)}{V(\phi)}\right]^2,\,\,\,
    \eta=\eta_V \equiv M^{*2}_{\mathrm{Pl}}\frac{V''(\phi)}{V(\phi)}.
\end{equation}
More generically, the slow-roll parameters $\epsilon,\eta$ are related by
\begin{eqnarray}\label{hubble flow par}
    \eta &&= \frac{\dot{\epsilon}}{H\epsilon}=\frac{2 \ddot{\phi}}{H\dot{\phi}}-\frac{2 \dot{H}}{H^{2}}= 2(\epsilon - \delta), \\
    \delta +\epsilon &&= -\frac{\ddot{\phi}}{H\dot{\phi}}-\frac{\dot{H}}{H^{2}}\simeq M^{*2}_{\rm Pl}\frac{V''}{V}=\eta_{V},
\end{eqnarray}
where $\delta \equiv -\ddot{\phi}/H\dot{\phi}$. The equation of motion of the scalar has been used, and the slow-roll approximation \eqref{slow-roll behavior} is used only once in the last step. We thus obtain the next-to-leading order slow-roll relation
\begin{equation}
    \eta = 4\epsilon - 2\eta_{V}. \label{NLOrelation}
\end{equation}

The inflationary period ends when $|\epsilon|$ reach unity, marking the transition to the reheating era~\cite{Bassett_2006}. A widely used measure to quantify the total amount of inflation is the number of $e$-foldings, defined by
\begin{equation}
    N \equiv \ln\frac{a_f}{a} = \int_{t}^{t_f} H{\rm d}t = \int_{\phi_f}^{\phi_i}\frac{\mathrm{d}\phi}{M^{*}_{\mathrm{Pl}}\sqrt{2\epsilon}}.
\end{equation}
At the leading order, $\epsilon \simeq \epsilon_V$ is used to simplify the calculation of $e$-folds. To improve the precision of $e$-folding number, we can include the higher order contribution of $\epsilon$ from $\eta$. From the expression for $\epsilon$ given in Eqn.~\eqref{hubble_slow-roll_parameter} with $\dot{H}$ in \eqref{eqn:hubble_rate}, $\dot{\phi}$ from \eqref{eqn:EoM_w_parameter} and \eqref{NLOrelation}, we find
\begin{equation}
    \epsilon = \frac{(\epsilon -3)^2 \epsilon_V}{(\eta_V-3 \epsilon -3)^2}.
\end{equation}
The asymptotic expansion of $\epsilon$ is therefore,
\begin{equation}\label{asymp_epsilon}
    \epsilon = \frac{9\epsilon_V}{(\eta_V - 3)^2} -\frac{54 (\eta_V-12) \epsilon_V^2}{(\eta_V-3)^5} + \mathcal{O}(\epsilon_V^3).
\end{equation}
This is the next-to-leading order expression for $\epsilon$ including $\eta$ contribution. In our work, we provide $e$-folding calculation at both the leading order and next-to-leading order slow-roll approximation with respect to the Hubble slow-roll parameter $\epsilon$:
\begin{align}
    N_1 &\simeq \int_{\phi_f}^{\phi_i}\frac{1}{M^{*}_{\mathrm{Pl}}\sqrt{2\epsilon_V}}\,\mathrm{d}\phi, \label{leading order N}\\
    N_2 &\simeq \int_{\phi_f}^{\phi_i}\frac{|\eta_V-3|} {3M^{*}_{\mathrm{Pl}}\sqrt{2\epsilon_V}}\,\mathrm{d}\phi, \label{next-to-leading order N}
\end{align}
where $\epsilon_V$ and $\eta_V$ are defined in Eqn~\eqref{slow-roll par}. $N_1$ is evaluated by the slow-roll approximation with only leading order, whereas $N_2$ is evaluated by the asymptotic expansion of $\epsilon$ up to next-to-leading order with leading $\eta_{V}$ contribution. 

\section{The Starobinsky inflationary model}\label{working1}
The Starobinsky model~\cite{STAROBINSKY198099} modifies the gravity action by using a generic function $f(R)$ which contains both linear Einstein-Hilbert action and the quadratic term $R^{2}$ originated from one-loop quantum gravity corrections. Incorporating quantum gravitational effects, the Starobinsky model predicts a graceful exit from inflation through the inclusion of an $R^2$ term~\cite{STAROBINSKY198099}. We can rewrite the action of the model into the traditional scalar-tensor theory given in Eqn.~\eqref{Ein_action} by determining the effective potential~(see Ref.~\cite{Burikham_2024} and references therein),
\begin{equation}\label{eff_potential}
    V(s) = \frac{Rf'(R)-f(R)}{2f'(R)^2},
\end{equation}
and using the field
\begin{equation}\label{sigma_phi}
    \sigma(\phi) \equiv \exp\left(\sqrt{\frac{2}{3}}\frac{\phi}{M_{\rm Pl}^*}\right) = f'(R),
\end{equation}
to describe the potential in a more transparent form for the measure construction.

\subsection{Determining the effective phase space}
The modified gravity function $f(R)$ for the Starobinsky model can be cast in the form $f(R) = R + \frac{\beta}{2}R^2$. By the effective potential~\eqref{eff_potential}, the corresponding potential can then be expressed as
\begin{equation}\label{potential_Star}
    V_{\rm Star}(\phi) = V_0\left(1-\frac{1}{\sigma(\phi)}\right)^2,
\end{equation}
where $V_0 = {M_{\rm Pl}^*}^2/4\beta$, and $\sigma(\phi)$ is given by Eqn.~\eqref{sigma_phi}. For $\beta$ value, we impose COBE normalization where the inflation vacuum energy at horizon exit can be estimated~\cite{Cheong:2020rao,Burikham_2024}~(see Ref.~\cite{Campista:2017ovq,SantosdaCosta:2020dyl} for more systematic analysis of the constraints),
\begin{equation}\label{COBE_normalization}
    \left.\frac{V}{\epsilon_V}\right|_{\phi_{\rm exit}} = 24\pi^2A_s{M^{*4}_{\rm Pl}}
    \simeq 0.027^4 {M^{*4}_{\rm Pl}}.
\end{equation}
The constraint correspondingly determines the $R^2$ coupling in terms of $e$-folds at horizon exit $N_e$ as follows,
\begin{equation}\label{beta_constraint}
    \beta \simeq \frac{N_e^2}{3(0.027)^4 {M^{*2}_{\rm Pl}}}.
\end{equation}

Next we define dimensionless coordinates for the effective phase space in Cartesian coordinates:
\begin{align}
    x &\equiv \frac{1}{2\sqrt{3}}\left(1-\frac{1}{\sigma} \right), \label{dimless coord.Star x}\\
    y &\equiv \sqrt{\frac{\beta}{6}}\frac{\dot{\phi}}{M^{*}_{\mathrm{Pl}}}. \label{dimless coord.Star y}
\end{align}
Since $\sigma(\phi)$ is the exponential function with a real-scalar field $\phi$, the restricted range of the coordinates $x$ is $(-\infty,1/2\sqrt{3})$. We also define polar coordinates $(z,\theta)$:
\begin{align}
    z &\equiv \sqrt{x^2+y^2} = \sqrt{\beta}H, \label{dimless corrd.Star z} \\
    \tan\theta &\equiv \frac{y}{x} = \frac{\sqrt{2\beta}\dot{\phi}\sigma}{M^{*}_{\mathrm{Pl}}(\sigma-1)}. \label{dimless coord.Star theta}
\end{align}
From $x\in(-\infty,1/2\sqrt{3})$,
a set of angles bounded at the Planck surface $H = M^{*}_{\mathrm{Pl}}$ is $(\theta_0,2\pi-\theta_0)$, where 
\begin{equation}\label{theta_0}
    \cos\theta_0 \equiv \frac{1}{2\sqrt{3}}\frac{1}{M^{*}_{\mathrm{Pl}}\sqrt{\beta}} \ll 1.
\end{equation}
We will use these dimensionless coordinates to plot trajectories in the effective phase space $(x,y)$.

Applying Eqn.~\eqref{eqn:motion} and Eqn.~\eqref{eqn:traj}, we obtain the vector field in $(x,y)$ coordinates $\textbf{v}= \dot{\textbf{x}}/\sqrt{\beta}$, where
\begin{equation}\label{vec_Cart_Star}
\begin{aligned}
    \dot{\mathbf{x}} = &\frac{y}{\sqrt{3}}\left(1-2\sqrt{3}x\right)\mathbf{\hat{x}} \\
    &- \left[3y\sqrt{x^2+y^2} + \frac{x}{\sqrt{3}} -2x^2\right]\mathbf{\hat{y}}.
\end{aligned}
\end{equation}
It can be written in polar coordinates $(z,\theta)$ as
\begin{equation}\label{vec_polar_Star}
\begin{aligned}
    \dot{\mathbf{x}} = & -3z^2\sin^2\theta\,\,\mathbf{\hat{z}}
    \\
    & -\left[ z^2\cos\theta\left(3\sin\theta-2\right) + \frac{z}{\sqrt{3}}\right]\boldsymbol{\hat{\theta}},
\end{aligned}
\end{equation}
where $x=z\cos\theta$, $y=z\sin\theta$, and the transformation
\begin{equation}
\left(\begin{array}{c}
\mathbf{\hat{z}}\\
\boldsymbol{\hat{\theta}}
\end{array}\right)=\left(\begin{array}{cc}
\cos\theta & \sin\theta\\
-\sin\theta & \cos\theta
\end{array}\right)\left(\begin{array}{c}
\mathbf{\hat{x}}\\
\mathbf{\hat{y}}
\end{array}\right).\label{coordrotation}
\end{equation}

\begin{figure*}
\begin{centering}
    \includegraphics[width=0.31\linewidth]{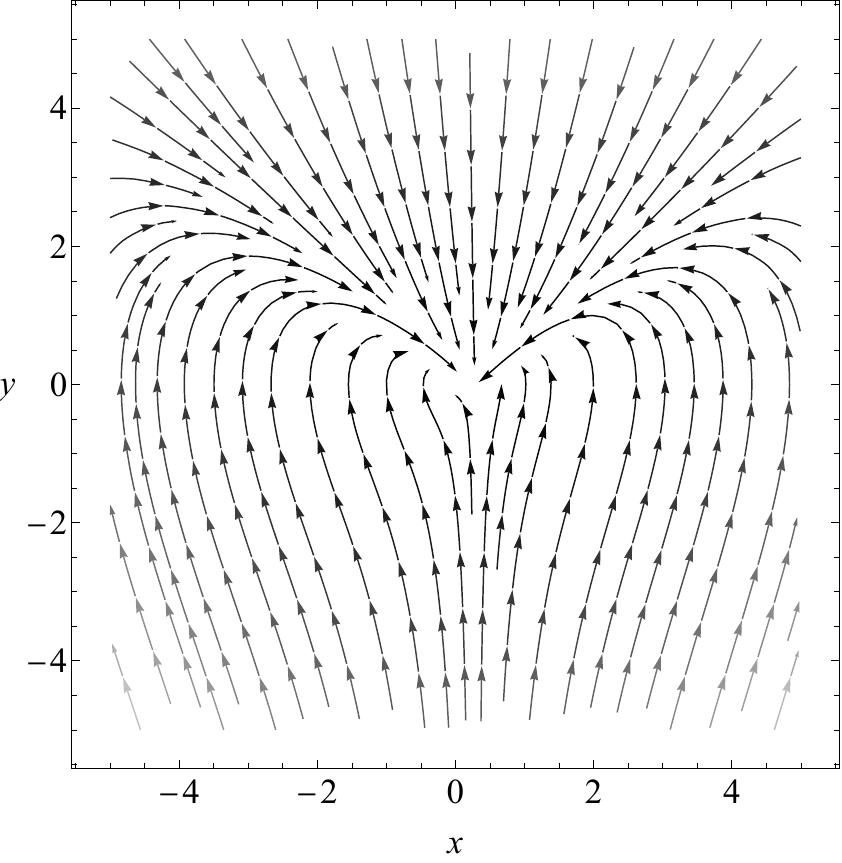}
    $\;\;\;\;$
    \includegraphics[width=0.31\linewidth]{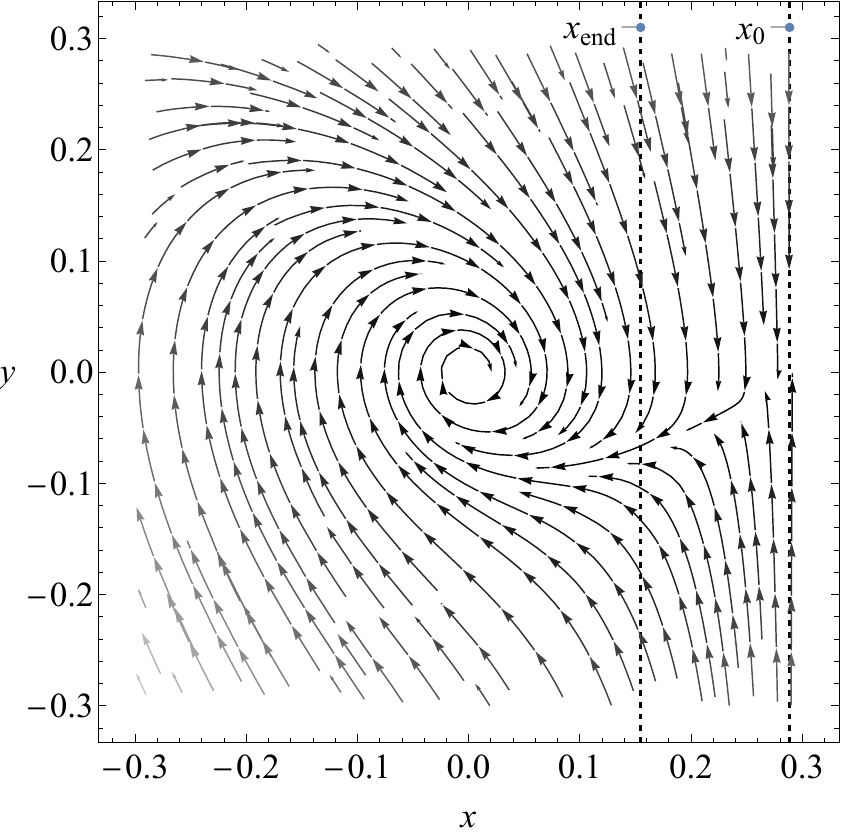}
    $\;\;\;\;$
    \includegraphics[width=0.31\linewidth]{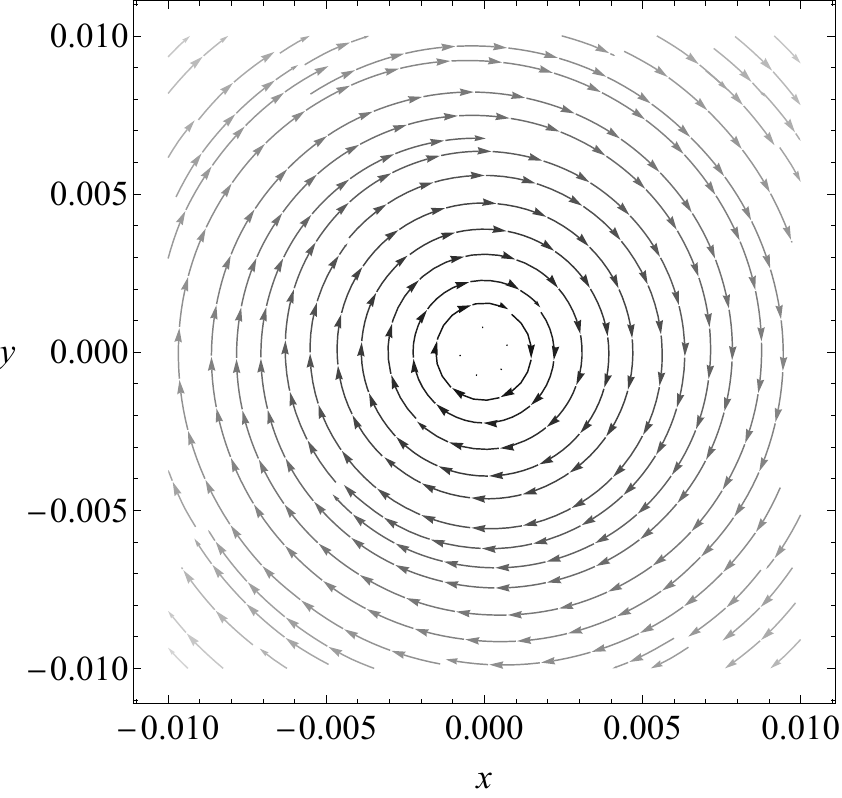}   
\end{centering}
    \caption{The flow vector fields in the effective phase space given by Eqn.~\eqref{vec_Cart_Star} with $M^{*}_\mathrm{Pl} = 1$, and $\beta = 0.1$. (Left) At the large effective field values, $x,y \gg 1$, there are apparent attractors, where the flow of trajectories is dense at $\theta = \arcsin(2/3)$~(both two angles). (Middle) We found an apparent attractor tracing down and turning into a curve around $\theta \simeq 0$ for $x>0$. For slow-roll region, $x$ is required to be within $(x_\mathrm{end}, x_0)$, corresponding to $\cos\theta_\mathrm{end}$ and $\cos\theta_0$ at the Planck surface respectively. (Right) For the small effective field value, $x,y \ll 1$, all trajectories converge to oscillatory solution.}
    \label{fig:vec_field_Star}
\end{figure*}

Using the flow vector field in Eqn.~\eqref{vec_Cart_Star}, we can plot the flow of trajectories in the effective phase space of $x-y$ plane as shown in Fig.~\ref{fig:vec_field_Star}. There are three apparent attractors: One appears at early universe ($H \gg 1$) in the large-field region at fixed angles, another oscillatory attractor appears at late universe ($H \ll 1$) in the small-field region, and the slow-roll attractor connecting between the large-field and the small-field regions. 

In particular, the apparent slow-roll attractor solution appears at $y \approx 0$ with $x > 0$, and then the flow changes into oscillatory attractor solution. For large field, the fixed angle attractor appears at $\theta = \arcsin(2/3)$. The final attractor at late universe appears as an oscillation in the small-field region, representing reheating era. By Liouville measure, the apparent attractor behavior where the flow of trajectories is dense implies that the measure density increase around this region. In this work, only the slow-roll and reheating attractors are physically relevant. The large field~($H>M^{*}_{\rm Pl}$) flow enters the quantum gravity realm and quantum measure is needed. 

\subsection{\texorpdfstring{Counting $e$-folding number in each trajectory}{Counting e-folds in each trajectory}}\label{counting_N_1}
For the Starobinsky potential~\eqref{potential_Star}, we compute the slow-roll parameters using Eqn.~\eqref{slow-roll par} to obtain
\begin{equation}\label{potential slow-roll par_Star}
    \epsilon_{V} = \frac{4}{3}\left(\frac{1}{\sigma -1}\right)^2, \quad
    \eta_{V} = -\frac{4\left(\sigma-2\right)}{3\left(\sigma-1\right)^2}.
\end{equation}
Inflation ends when $\epsilon=1$ and we stop counting the $e$-folding number at this point. From the expression for $\epsilon$ given by the leading-order slow-roll approximation, $\epsilon \simeq \epsilon_V$, and the next-to-leading-order asymptotic expansion in Eqn.~\eqref{asymp_epsilon}, we found the end points as given in Table~\ref{tab:inflation_end_point}.

\begin{table}[t]
    \caption{the end point of inflation $\sigma_{\rm end}$ when $\epsilon = 1$ from leading-order slow-roll approximation and the next-to-leading-order asymptotic expansion.}
\begin{ruledtabular}
\begin{tabular}{p{1.1in}|p{2in}}
    expression for $\epsilon$ & $\sigma_{\rm end}$ when $\epsilon = 1$ \\ \hline 
    $\begin{aligned}~\\[-2ex]
    \hspace{2.5em} \epsilon_V \hspace{15em} &  \hspace{-7em} 1+\frac{2}{\sqrt{3}} \\
    \frac{9\epsilon_V}{(\eta_V-3)^2}  \hspace{13em} & \hspace{-7em} \frac{1}{9}\left(3 \sqrt{3}+\sqrt{67-12 \sqrt{3}}+7\right) \\[1ex]
    \end{aligned}$ 
\end{tabular}
\end{ruledtabular}
\label{tab:inflation_end_point}
\end{table}

The number of $e$-folds can be calculated by using both leading-order and next-to-leading-order expression of $\epsilon$ from Eqn~\eqref{leading order N} and~\eqref{next-to-leading order N}, respectively. The final inflaton value $\phi_f$ can also be determined by $\phi_f = \sqrt{3/2}\,\ln\sigma_{\rm end}$. Using Eqn~\eqref{potential slow-roll par_Star}, we obtain the $e$-folding number achieved by the initial inflaton $\phi$ as follows:
\begin{align}
    N_1 &= \frac{3}{4}\left[\sigma(\phi)-\ln\left(\frac{\sqrt{3}}{\sqrt{3}+2}\sigma(\phi)\right)\right] + A_0, 
    \label{leading order e-folding function}\\
    N_2 &= \frac{1}{12}\biggl[9\sigma(\phi)-\ln\left(\left(\sigma(\phi)-1\right)^4\sigma(\phi)\right)\Biggr]+B_0,
    \label{next-to-leading order e-folding function}
\end{align}
where 
\begin{align}
    A_0 &= -\frac{3}{4}\left[1+\frac{2}{\sqrt{3}}\right] \approx -1.616, \label{const_e-folds_LO}\\
    B_0 &= \frac{1}{12}\Biggl[4\ln \left(\frac{1}{9}\left(3\sqrt{3}+\sqrt{67-12 \sqrt{3}}-2\right)\right) \nonumber \\ 
    &\mathrel{\phantom{=}} +\ln\left(\frac{1}{9} \left(3 \sqrt{3}+\sqrt{67-12 \sqrt{3}}+7\right)\right) \nonumber \\ 
    &\mathrel{\phantom{=}} -3\sqrt{3}-\sqrt{67-12 \sqrt{3}}-7 \Biggr] \approx -1.486. \label{const_e-folds_NLO}
\end{align}

\begin{figure}[t]
    \centering
    \includegraphics[width=\linewidth]{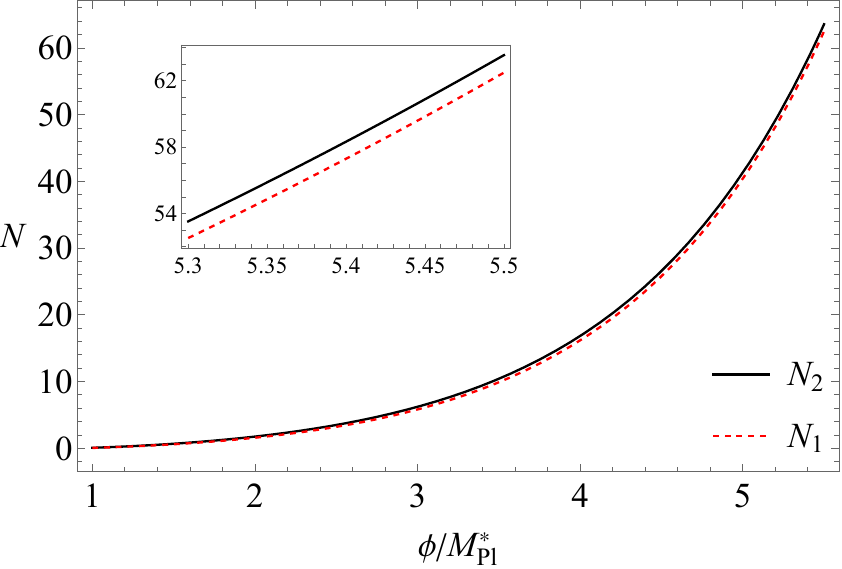}
    \caption{The $e$-folding number achieved by the initial inflaton $\phi$. As shown in the zoom-in figure, $N_2$, evaluated using Eqn.\eqref{asymp_epsilon}, results in a greater number of $e$-folds compared to $N_1$, which is calculated using Eqn.~\eqref{slow-roll par}. $N_2$ includes contributions from both $\epsilon_V$ and $\eta_V$, whereas $N_1$ considers only $\epsilon_V$.}
    \label{fig:e-folds_function_LO-NLO}
\end{figure}

\begin{figure}[t]
\begin{center}
    \includegraphics[width=\linewidth]{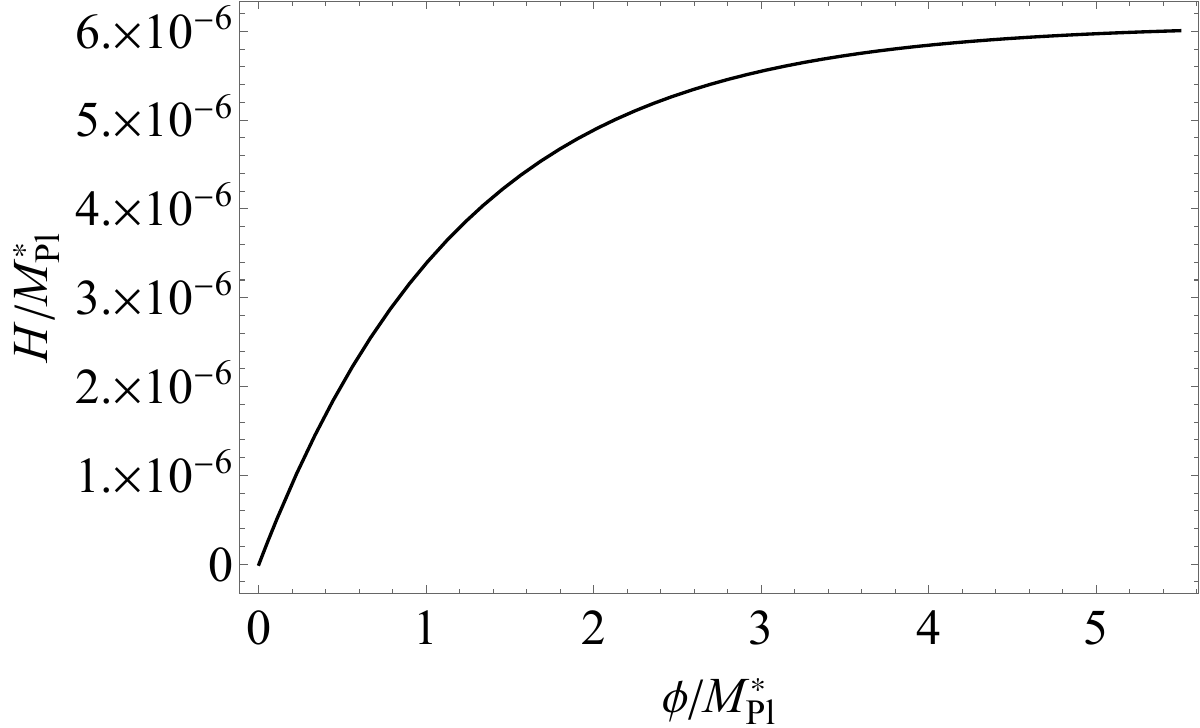}
\end{center}
    \caption{Hubble parameter versus inflaton value under slow-roll condition, the Hubble parameter is always sub-Planckian throughout the slow-roll inflation.} 
    \label{fig:Hubble_SR_vs_phi}
\end{figure} 
Note that $N_{1,2}$ diverge as $\phi_{\rm UV}\to \infty~(\sigma\to\infty)$, i.e., Starobinsky potential has infinitely asymptotically flat plateau and a cutoff is required.

Conveniently, our aim is to express the number of $e$-folds for each trajectory in terms of its coordinates $\theta$ on the Planck surface, rather than the coordinate $x$ when it enters the slow-roll regime. Consequently, the total number of $e$-folds achieved by a trajectory starting at angle $\theta$ on the Planck surface can be evaluated from Eqn.~\eqref{leading order e-folding function} and Eqn.~\eqref{next-to-leading order e-folding function} by using an expression for $\sigma$ in terms of $\theta$. According to the coordinates transformation, $x = M_{\rm Pl}^*\sqrt{\beta}\cos\theta$, we combine Eqn.~\eqref{dimless coord.Star x} with the definition of $\cos\theta_0$ in Eqn.~\eqref{theta_0} to obtain
\begin{equation}\label{sigma_theta}
    \sigma(\theta) = \frac{1}{1-\frac{\cos\theta}{\cos\theta_0}} = \frac{1}{1-u},
\end{equation}
where $u \equiv \cos\theta/\cos\theta_0$.

The slow-roll regime $\epsilon_V, |\eta_V| < 1$ occurs in the region where $\phi > 0$, and $0<x<1/2\sqrt{3}$, as given by Eqs.~\eqref{sigma_phi} and~\eqref{dimless coord.Star x}. Since Starobinsky inflation starts from the positive scalaron side of the potential ($\phi > 0$) where it indefinitely flattens as $\phi \to \infty$ and $\epsilon \simeq \epsilon_V \to 0$, we can achieve infinite $e$-folds when $x = 1/(2\sqrt{3})$ or $\theta = \theta_0$. Namely, the $e$-folding number in both Eqn.~\eqref{leading order e-folding function} and Eqn.~\eqref{next-to-leading order e-folding function} exhibits a singularity at $u = 1$. However, inflation should actually start at certain high-energy scale where the number of $e$-folds is regulated by a UV~(ultraviolet) cutoff $u_{\mathrm{UV}}$, presumably before the quantum gravity scale. From Eqn.~\eqref{dimless coord.Star x}, the end point of the inflation is at $x_{\rm end} = (2-\sqrt{3})/\sqrt{3}$. Therefore, the physical range of $x$ that satisfies the slow-roll condition is $x \in \left(x_{\rm end}, u_{\rm UV}/(2\sqrt{3})\right]$ and the corresponding range of angle $\theta$ is $\theta \in [\theta_{\mathrm{UV}},\theta_{\rm end})$,
where
\begin{align}
    \cos\theta_{\mathrm{UV}} &= \frac{u_{\mathrm{UV}}}{2\sqrt{3}M^{*}_{\mathrm{Pl}}\sqrt{\beta}}, \label{theta_cutoff} \\
    \cos\theta_{\rm end} &= \frac{2-\sqrt{3}}{\sqrt{3}M^{*}_{\mathrm{Pl}}\sqrt{\beta}}.\label{theta_end}
\end{align}

Using cosmological observations to constrain the values of inflaton cutoff $\phi_{\rm UV}$, we consider two key observables: the scalar spectral index $n_s$ and the tensor-to-scalar ratio $r$~(following Ref.~\cite{Ivanov_2022}). These quantities can be related to potential slow-roll parameters in inflationary models, as shown in the following expressions~\cite{Liddle_1994,Lyth_Liddle_1992}, 
\begin{align}
    \label{n_s with slow-roll parameter} n_s &= 1 - 6\epsilon_V + 2\eta_V, \\
    \label{r with slow-roll parameter} r &= 16\epsilon_V.
\end{align}
These parameters are also constrained by cosmological observation as~\cite{Planck:2018jri,BICEP:2021xfz}
\begin{align}
    \label{n_s constraint} n_s = 0.9649 \pm 0.0042 \;\;(68\% \;{\rm CL}), \\
    \label{r constraint} r < 0.036 \;\;(95\%\;{\rm CL}).
\end{align}
Taking $\epsilon_V$ and $\eta_V$ by Eqn.~\eqref{potential slow-roll par_Star}, we obtain the constraint of inflaton cutoff $\phi_{\rm UV}$ as shown in Table~\ref{tab: n_s constraints Star}. This restricted range also satisfies the weaker constraint on $r$. We will use this range of inflaton cutoff, $5.217 < \phi_{\rm UV}/M_{\rm Pl}^* < 5.501$, in subsequent sections.

\begin{table}[t]
    \centering
    \caption{Observational constraints on the scalar spectral index $n_s$ requires $5.217 < \phi_{\rm UV}/M_{\rm Pl}^* < 5.501$. The corresponding value of $r$ is then computed and listed in the table. The $e$-folding number at $\phi_{\rm UV}$ is calculated by Eqn.~\eqref{leading order e-folding function}, and~\eqref{next-to-leading order e-folding function}.}
    $\begin{array}{|c|c|c|}
        \hline
        n_s & 0.9607 & 0.9691 \\ \hline
         \phi_{\rm UV}/M_{\rm Pl}^* & 5.217 & 5.501  \\ \hline
         r & 0.0044 & 0.0027\\ \hline
         N_1(\phi_{\rm UV}) & 48.86 & 62.54 \\ \hline
         N_2(\phi_{\rm UV}) & 49.83 & 63.59 \\ \hline
    \end{array}$
    \label{tab: n_s constraints Star}
\end{table}

\subsection{\texorpdfstring{The expectation value of $e$-folding number and the probability function}{The expectation value of e-folds and the probability function}}\label{sec:exp_e_folds_Star}
From previous section, we can compute the $e$-folds in a given trajectory that starts with the initial inflaton $\phi$ at angular coordinate $\theta$ of the Planck surface given by Eqn.~\eqref{leading order e-folding function} and~\eqref{next-to-leading order e-folding function} at the leading order and next-to-leading order. Now we compute the expectation value of $e$-folding number in the Starobinsky model under the conserved measure. 

First we find the measure $\omega$ on the effective phase space in order to construct the probability distribution function. In the $(z,\theta)$ coordinates, we can write Eqn.~\eqref{vec_polar_Star} in the early universe scale ($z$ is very large) and the dimensionless velocity in polar coordinates takes the following form, 
\begin{equation}
    \mathbf{v} \simeq -\frac{3z^2\sin^2\theta}{\sqrt{\beta}}\,\,\mathbf{\hat{z}}
    -\frac{z^2\cos\theta}{\sqrt{\beta}}\left(3\sin\theta-2\right)\boldsymbol{\hat{\theta}}.
\end{equation}
The requirement for the measure $\omega$ in Eqn.~\eqref{conservation} to be conserved under time evolution can be expressed as
\begin{equation}\label{pde_Star}
\begin{aligned}
    3z\partial_z\omega &= \cot\theta\left(2\csc\theta-3\right)\partial_\theta\omega \\
    &\mathrel{\phantom{=}} -\left(3\cot^2\theta+2\csc\theta-6\right)\omega.
\end{aligned}
\end{equation}
To find the solution for $\omega$ in Eqn.~\eqref{pde_Star}, we use the separation of variables:
\begin{equation}
    \omega\left(z,\theta\right) = R(z)\Theta(\theta),
\end{equation}
substitute into Eqn.~\eqref{pde_Star} to obtain
\begin{align}
    3z\frac{\partial_z R}{R}+9 = \frac{\partial_\theta\left[\left(2\cos\theta-3\sin\theta\cos\theta\right)\Theta\right]}{\Theta\sin^2\theta}.
\end{align}
The solutions are
\begin{equation}\label{radial_sol_Star}
    R(z) = Cz^{(m-9)/3}
\end{equation}
and
\begin{equation}\label{aungular_sol_Star}
\begin{aligned}
    \Theta(\theta) &= C(2-3\sin\theta)^{-(1+4m/15)} \\
    &\mathrel{\phantom{=}}\times (1-\sin\theta)^{(m-1)/2} \\
    &\mathrel{\phantom{=}}\times (1+\sin\theta)^{(m-5)/10},
\end{aligned}
\end{equation}
where $m \in \mathbb{R}$ is arbitrary. Since the measure plays a role of the probability distribution on the effective phase space, we require the measure to be positive everywhere. Therefore, the general solution for the measure $\omega$ takes the form
\begin{equation}
    \omega = \sum_m C_m z^{\frac{m-9}{3}} \left|
    \frac{(1-\sin\theta)^{\frac{m-1}{2}}(1+\sin\theta)^{\frac{m-5}{10}}}
    {(2-3\sin\theta)^{\left(1+\frac{4m}{15}\right)}}
    \right|.
\end{equation}
Moreover, the physical solution should satisfy the following conditions~\cite{Remmen_2014}:
\begin{enumerate}
    \item $\omega$ is finite everywhere except on the apparent attractor trajectory.
    \item At fixed $\theta$, trajectories converge and become more squeezed together as the radial component $z$ decreases. It implies that $\omega$ is inversely proportional to $z$, so $m \leq 9$.
    \item The measure $\omega$ is infinitely differentiable everywhere except the apparent attractor solution, there is only one case that satisfies this condition which is $m=5$.
\end{enumerate}
From these conditions, we then select $m = 5$ as the physical solution. The measure then becomes
\begin{equation}\label{omega_Star}
    \omega(\theta) \propto \frac{1}{z^{4/3}}
    \frac{(1-\sin\theta)^{2}}
    {\left|2-3\sin\theta\right|^{7/3}}.
\end{equation}

The next step is to construct the probability distribution over the space of trajectories, parametrized by $\theta$ on a surface of constant $H$, in this case, we choose the Planck surface ($H = M^{*}_{\mathrm{Pl}}$). The probability distribution takes a form $\omega(H,\theta)|\dot{H}|$ where $\dot{H}$ is defined in Eqn.~\eqref{eqn:hubble_rate}:
\begin{equation}\label{hubble_rate_w_polar_coord}
    \dot{H} = -\frac{\dot{\phi}^2}{2M^{*2}_{\mathrm{Pl}}} = -\frac{3}{\beta}z^2\sin^2\theta.
\end{equation}
Therefore, the probability distribution on the space of trajectories on the Planck surface, where $z = \sqrt{\beta} M^{*}_\mathrm{Pl}$, is given by
\begin{equation}\label{prob_dist_Star}
    P(\theta) = C\frac{(1-\sin\theta)^{2}}
    {\left|2-3\sin\theta\right|^{7/3}}\sin^2\theta,
\end{equation}
where $C$ is a normalization factor. 

In order to determine the normalization factor, we focus exclusively on the region within the UV cutoff that satisfies the slow-roll conditions, specifically for $\theta \in (\theta_\mathrm{UV},\theta_{\rm end})$, where $\theta_\mathrm{UV}$ and $\theta_{\rm end}$ are defined in Eqn.~\eqref{theta_cutoff} and Eqn.~\eqref{theta_end}, respectively. By imposing $\cos\theta_0$, $\cos\theta_{\rm end} \ll 1$, we can simplify the probability distribution $P(\theta)\,\mathrm{d}\theta$. Specifically, we approximate $\sin\theta \simeq 1-\frac{1}{2}\cos^2\theta$ in the interval $(\theta_0, \pi/2)$, and $\sin\theta \simeq \frac{1}{2}\cos^2\theta -1$ in the interval $(3\pi/2,2\pi-\theta_0)$. Thus, the requirement that the total probability equals unity, given by the normalization condition $\int P(\theta)\,\mathrm{d}\theta = 1$ for the probability distribution~\eqref{prob_dist_Star}, simplifies to the following expression:
\begin{align}\label{normalization}
    1 &= \int_{\theta_{\rm end}}^{\theta_{\rm UV}} \frac{C}{4}\cos^4\theta\,\mathrm{d}\cos\theta \nonumber \\
    &\mathrel{\phantom{=}} -\int_{2\pi-\theta_{\rm end}}^{2\pi-\theta_\mathrm{UV}} \frac{4C}{5^{7/3}}\left(\frac{3}{10}\cos^2\theta -1\right)\,\mathrm{d}\cos\theta,  
\end{align}\\
over the slow-roll region on the Planck surface. Applying the identity $\cos\theta = \cos(2\pi-\theta)$, we can express the probability distribution in terms of $u = \cos\theta/\cos\theta_0$ as follows:
\begin{align}\label{approx_prob_dist_Star}
    P(u) = C\left[\frac{1}{4}u^4\cos^4\theta_0 -\frac{4}{5^{7/3}}\left(\frac{3}{10}u^2\cos^2\theta_0-1\right)\right].
\end{align}
The normalization condition then takes the form $\int P(u)\,\mathrm{d}u = 1$ over the region $(u_{\rm end}, u_{\rm UV})$ and the normalization factor $C$ is given by
\begin{align}\label{norm_factor}
    C^{-1} &= \frac{1}{20}\left(u^5_\mathrm{UV} - u^5_{\rm end}\right)\cos^4\theta_0 \nonumber \\
            &\mathrel{\phantom{=}} - \frac{4}{5^{7/3}}\frac{1}{10}\left(u^3_\mathrm{UV} - u^3_{\rm end}\right)\cos^2\theta_0 \nonumber \\
            &\mathrel{\phantom{=}} + \frac{4}{5^{7/3}}\left(u_\mathrm{UV} - u_{\rm end}\right),
\end{align}
where $u_{\rm UV}$ depends on the chosen setting, while $u_{\rm end} = \cos\theta_{\rm end}/\cos\theta_0$ depends on $e$-folding function used. Specifically, the leading-order expression~\eqref{leading order e-folding function} arises from the slow-roll approximation, while the next-to-leading-order expression~\eqref{next-to-leading order e-folding function} come from the asymptotic expansion. According to Table~\ref{tab:inflation_end_point} and the expression~\eqref{sigma_theta}, we get
\begin{align}
    u_{\rm end, 1} &= 4-2\sqrt{3} \approx 0.536,\\
    u_{\rm end, 2} &= 1-\frac{9}{7+3\sqrt{3}+\sqrt{67-12 \sqrt{3}}} \approx 0.526,
\end{align}
for $N_1$ and $N_2$, respectively. Note that this $P(u)$ is the probability that we find trajectories which intersect Planck surface at any $N$ $e$-folds.

\begin{figure}[t]
    \centering
    \includegraphics[width=\linewidth]{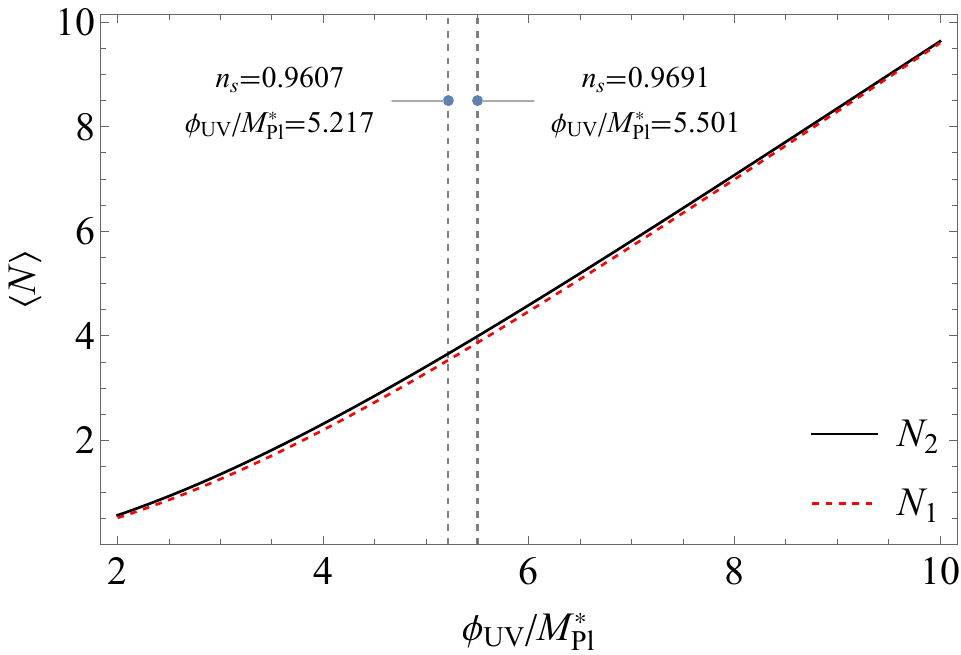}
    \caption{Expectation value of $e$-folds $\langle N \rangle$ from the leading-order expression~\eqref{leading order e-folding function} and the next-to-leading-order expression~\eqref{next-to-leading order e-folding function}, using the conserved measure~\eqref{approx_prob_dist_Star} for the Starobinsky model versus $\phi_\mathrm{UV}$ cutoff.} 
    \label{fig:e-folds_vs_cutoff}
\end{figure}

Finally, we can compute the expectation value of $e$-folds attained in Starobinsky inflationary model from all trajectories in the effective phase space starting at angle $\theta$ on the Planck surface. Since the number of $e$-folds can be approximate by the slow-roll formula Eqn.~\eqref{leading order e-folding function} and \eqref{next-to-leading order e-folding function}, therefore we calculate $\langle N \rangle = \int N(u)P(u)\,\mathrm{d}u$ over the slow-roll region $(u_{\rm end}, u_{\rm UV})$.  By using the probability distribution~\eqref{approx_prob_dist_Star} with the normalization factor~\eqref{norm_factor}, we obtain the explicit expressions for $\langle N \rangle$ as follows:
\begin{widetext}
\begin{align}
    \label{N_average_LO} \langle N_1 \rangle &= \left[D(u_{\rm UV}-u_{\rm end,1})-\frac{D}{10}\left(u^3_{\rm UV} - u^3_{\rm end,1}\right)\cos^2\theta_0+\frac{1}{20}\left(u^5_{\rm UV} - u^5_{\rm end,1}\right)\cos^4\theta_0\right]^{-1}\nonumber \\
    &\mathrel{\phantom{=}}\times \biggl[\left(-\frac{3D}{4}+\frac{3D\cos^2\theta_0}{10}-\frac{9\cos^4\theta_0}{40}\right)u_{\rm UV} +\left(\frac{3D\cos^2\theta_0}{20}-\frac{9\cos^4\theta_0}{80}\right)u^2_{\rm UV} \nonumber \\
    &\mathrel{\phantom{=}} +\left(\frac{D\cos^2\theta_0}{40}-\frac{3\cos^4\theta_0}{40}\right)u^3_{\rm UV} -\frac{9\cos^4\theta_0}{160}u^4_{\rm UV} -\frac{3\cos^4\theta_0}{400}u^5_{\rm UV} \nonumber \\
    &\mathrel{\phantom{=}} -\left(\frac{3D}{2}-\frac{3D\cos^2\theta_0}{10} +\frac{9\cos^4\theta_0}{40}\right)\ln(1-u_{\rm UV}) \nonumber \\ 
    &\mathrel{\phantom{=}} -\left(\frac{3D}{4}-\frac{3D\cos^2\theta_0}{40}u^2_{\rm UV}-\frac{3\cos^4\theta_0}{80}u^4_{\rm UV}\right)u_{\rm UV}\ln\left(\frac{2\sqrt{3}-3}{1-u_{\rm UV}}\right) +A_1 \biggr] +A_0,
\end{align}
\end{widetext}
\begin{widetext}
\begin{align}
    \label{N_average_NLO}\langle N_2 \rangle &= \left[D(u_{\rm UV}-u_{\rm end,2})-\frac{D}{10}\left(u^3_{\rm UV} - u^3_{\rm end,2}\right)\cos^2\theta_0+\frac{1}{20}\left(u^5_{\rm UV} - u^5_{\rm end,2}\right)\cos^4\theta_0\right]^{-1}\nonumber \\
    &\mathrel{\phantom{=}} \times \biggl[\left( -\frac{D}{12} + \frac{4D\cos^2\theta_0}{15} -\frac{5\cos^4\theta_0}{24}\right)u_{\rm UV} + \left(\frac{2D\cos^2\theta_0}{15}-\frac{5\cos^4\theta_0}{48}\right)u^2_{\rm UV} \nonumber \\
    &\mathrel{\phantom{=}} +\left(\frac{D\cos^2\theta_0}{360}-\frac{5\cos^4\theta_0}{72}\right)u^3_{\rm UV} - \frac{5\cos^4\theta_0}{96}u^4_{\rm UV}-\frac{\cos^4\theta_0}{1200}u^5_{\rm UV} \nonumber \\
    &\mathrel{\phantom{=}} + \left(\frac{5D}{12}u_{\rm UV}-\frac{D\cos^2\theta_0}{24}u^3_{\rm UV}+\frac{\cos^4\theta_0}{48}u^5_{\rm UV}-\frac{7D}{6}+\frac{4D\cos^2\theta_0}{15}-\frac{5\cos^4\theta_0}{24}\right)\ln(1-u_{\rm UV}) \nonumber \\
    &\mathrel{\phantom{=}} -\left(\frac{D}{3}-\frac{D\cos^2\theta_0}{30}u^2_{\rm UV}+\frac{\cos^4\theta_0}{60}u^4_{\rm UV}\right)u_{\rm UV}\ln (u_{\rm UV}) +B_1 \biggr] + B_0,
\end{align}
\end{widetext}
where $D=4/5^{7/3}$, $A_1 \approx 0.070$, $B_1 \approx 0.073$, $A_0 \approx -1.616$, $B_0 \approx -1.149$, $u_{\rm end,1} \approx 0.536$, and $u_{\rm end,2} \approx 0.526$.
Finally, we set $N_e = 60$ for the value of $\beta$ given by Eqn.~\eqref{beta_constraint}. Figure~\ref{fig:e-folds_vs_cutoff} shows $\langle N_1 \rangle$ and $\langle N_2 \rangle$ plotted as functions of $\phi_{\rm UV}$ cutoff, where $\phi_{\rm UV}$ and $u_{\rm UV}$ is related by
\begin{equation}\label{u_phi}
    u = 1-\exp(-\sqrt{\frac{2}{3}}\frac{\phi}{M^{*}_\mathrm{Pl}}).
\end{equation}
However, $u_{\rm UV}$, determined by $\phi_{\rm UV}$, represents the upper limit of the initial value of the inflaton field and must be consistent with the cosmological observables. We then take $\phi_{\rm UV}\in [5.22,5.50]M^{*}_{\rm Pl}$, which satisfies the constraint on the spectral index, $n_s = 0.9658 \pm 0.0040\,\,\,(68\%\,\,{\rm CL})$, for the Starobinsky inflationary model. In this allowed range of the inflaton cutoff, the expectation value of $e$-folds and the $e$-folding number with the cutoff $\phi_{\rm UV}$, obtained from the leading-order expression and the next-to-leading-order expression are given in Table~\ref{tab:N_average_vs_phi_UV}.
\begin{table}[t]
    \caption{The expectation value of $e$-folding number and the $e$-folding number in the Starobinsky model, approximately computed using the leading-order expression~\eqref{leading order N} and the next-to-leading-order expression~\eqref{next-to-leading order N}, for different values of cutoff $\phi_{\rm UV}$.}
    \centering
    $$\begin{array}{||c|c|c|c|c||}
    \hline
    \phi_{\rm UV} & \langle N_1 \rangle & \langle N_2 \rangle & N_1(\phi_{\rm UV}) & N_2(\phi_{\rm UV})\\[0.5ex] \hline \hline
         5.221 & 3.541 & 3.665 & 49.02 & \textbf{50.00}  \\
         5.244 & 3.567 & 3.691 & \textbf{50.00} & 50.99  \\
         5.300 & 3.633 & 3.757 & 52.53 & 53.53      \\
         5.400 & 3.751 & 3.874 & 57.30 & 58.33      \\
         5.433 & 3.789 & 3.913 & 58.96 & \textbf{60.00}  \\
         5.453 & 3.813 & 3.967 & \textbf{60.00} & 61.04  \\
         5.500 & 3.869 & 3.992 & 62.48 & 63.54      \\ \hline
    \end{array}$$
    \label{tab:N_average_vs_phi_UV}
\end{table}

In computing the $e$-folds average, we initially employed an approximate form of the probability distribution, which simplified the calculation and allowed for analytical integration. This approach provided an efficient way to estimate the $e$-folds average. However, to verify the accuracy of this approximation, we subsequently performed a more rigorous calculation by using the full version of the probability distribution in a numerical integration. The resulting discrepancy between the two approaches was on the order of $10^{-15}$, indicating that the approximation method is highly accurate for our propose, with only a negligible difference from the full calculation.

\begin{figure}[t]
\begin{center}
    \includegraphics[width=\linewidth]{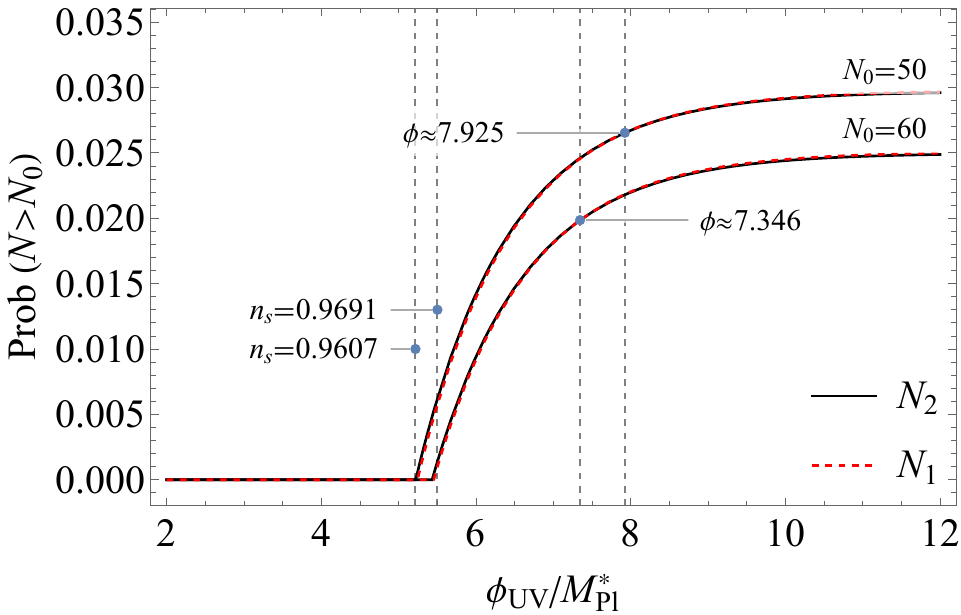}
\end{center}
    \caption{The probability to obtain at least 50 and 60 $e$-folds as functions of $\phi_{\rm UV}$, computed by using the conserved measure~\eqref{approx_prob_dist_Star} on the $H = M^{*}_{\rm Pl}$ surface. At $\phi_{\rm UV} \approx 7.925 M^{*}_{\rm Pl}$ for $N_0 =50$ and $\phi_{\rm UV} \approx 7.346 M^{*}_{\rm Pl}$ for $N_0 = 60$, the probability comparison indicates that ${\rm Prob}(N_2>60) \ge {\rm Prob}(N_1>60)$ for inflaton value lower than these thresholds, while ${\rm Prob}(N_2>60) < {\rm Prob}(N_1>60)$ for inflaton values above the thresholds. However, the differences are insignificantly small. 
    } 
    \label{fig:fractionU_vs_cutoff}
\end{figure}

Additionally, the probability that the total $e$-folding number $N$ exceed $N_0$ is 
\begin{equation}\label{fractionU_Star}
    \mathrm{Prob}(N > N_0) = \int_{u_{N=N_0}}^{u_{\rm UV}} P(u)\,\mathrm{d}u,
\end{equation}
where the total $e$-folding number $N$ can be computed using $N_1$ in Eqn.~\eqref{leading order e-folding function} and $N_2$ in Eqn.~\eqref{next-to-leading order e-folding function}. For sufficiently large $e$-folds to solve the horizon problem, we consider the benchmark $N_0 = 50, 60$ as shown in Fig.~\ref{fig:fractionU_vs_cutoff}. For a trajectory intersecting the Planck surface to reach at least 50 and 60 $e$-folds within the effective phase space, the inflaton field must reach certain threshold values given in Table~\ref{tab:N_average_vs_phi_UV}. 

\begin{figure}[t]
    \centering
    \includegraphics[width=\linewidth]{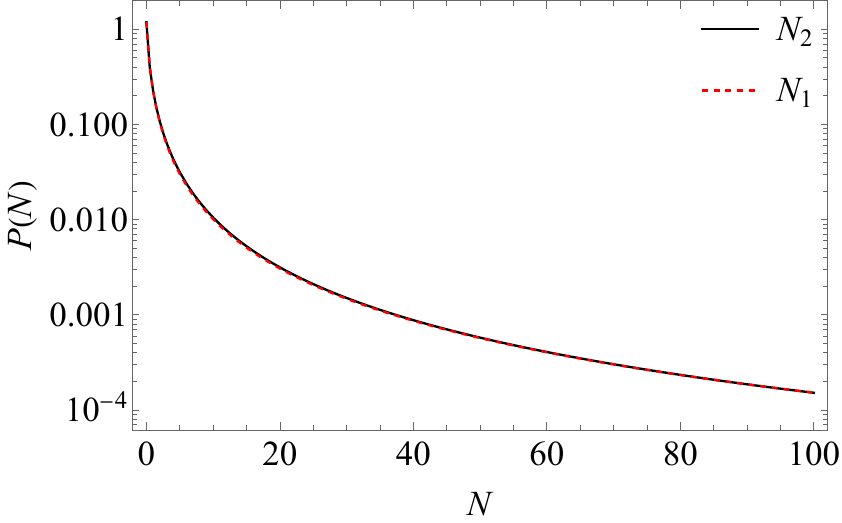}
    \caption{The probability distribution on the space of trajectories versus the $e$-folding number $N$ with inflaton cutoff at $\phi_{\rm UV} = 5.5 M^{*}_{\rm Pl}$.}
    \label{fig:P(N)_vs_N_Star}
\end{figure}

For a specific inflaton cutoff of $\phi_{\rm UV} = 5.5 M^{*}_{\rm Pl}$, the probability distributions of trajectory space as a function of the $e$-folding number for $0 \leq N_1,N_2 \leq N_{\rm max} = N(\phi_{\rm UV})$---are numerically plotted in Fig.~\ref{fig:P(N)_vs_N_Star}. The probability distribution decreases with increasing $N$ and approaches zero in the limit $N \to \infty$. Consequently, the fraction of universes achieving more than 50 and 60 $e$-folds saturates at 0.0297 and 0.0250 respectively, as shown in Fig.~\ref{fig:fractionU_vs_cutoff}.

\section{The extended Starobinsky Inflationary Model}\label{sec:working2}
\subsection{Determining the effective phase space}
\begin{figure}[t]
    \centering
    \includegraphics[width=\linewidth]{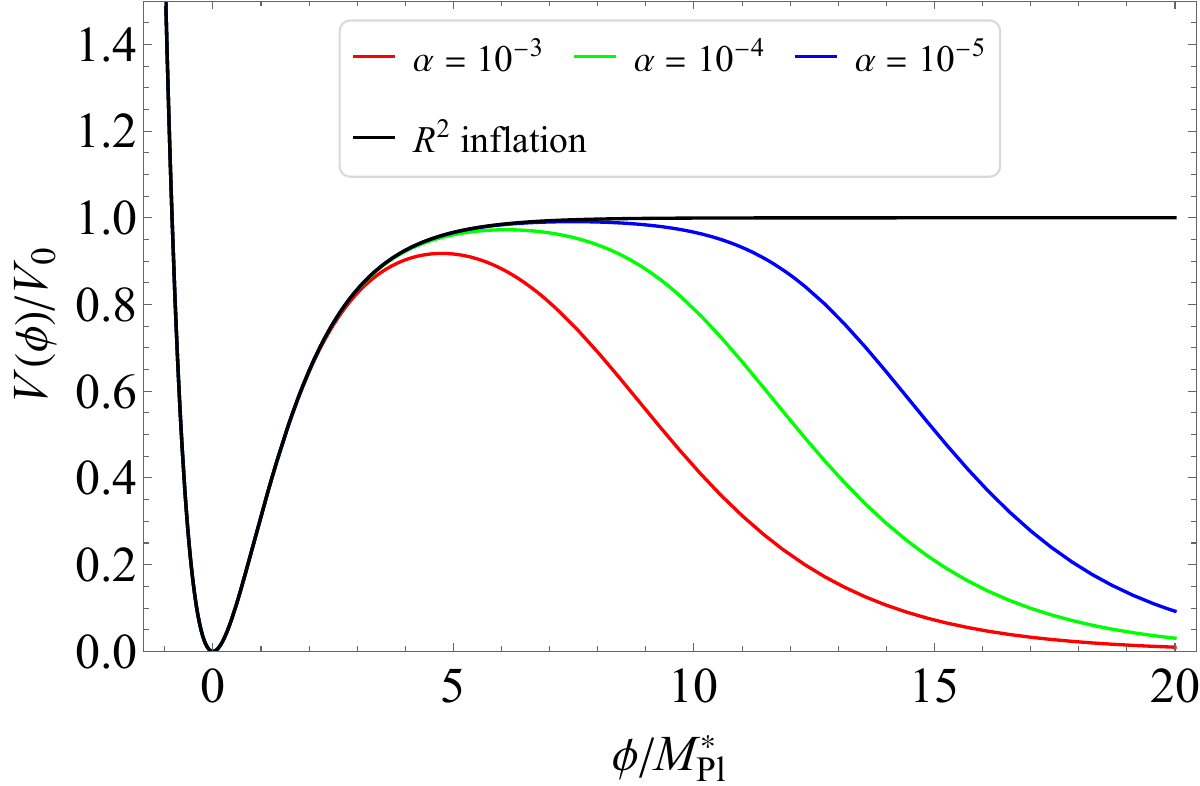}
    \caption{Potentials of the extended Starobinsky model for different values of parameter $\alpha$. Inflaton starts the slow-roll at $\phi_{\rm max}$ where the potential is maximum and slow-rolls to the left side with lower fields towards the quasi-stable mininum. The potential converges to the Starobinsky model, $R^2$ inflation, with infinite almost-flat plateau in the high fields as $\alpha \to 0$.}
    \label{fig:potential_vs_alpha}
\end{figure}
We now extend the Starobinsky model to include the higher-order terms in $R$. A simple modification is by adding an $R^3$ term as a phenomenological quantum gravity corrections, leading to a generic $\left(\displaystyle{R+\frac{\beta}{2}R^2+\frac{\gamma}{3}R^3}\right)$ gravity action~\cite{Cheong:2020rao}. The model satisfies the observational constraints for the scalar perturbation index $n_s$ as long as the $R^3$ term is kept relatively small comparing to the dominating $R^2$ term, so that $\gamma \ll \beta^2$~\cite{Cheong:2020rao,Burikham_2024,Ivanov_2022}. From Eqn.~\eqref{eff_potential}, the corresponding inflaton scalar potential for the extended Starobinsky inflationary model takes the form 
\begin{align}\label{potential_exStar}
    V_{\rm exStar}(\phi) &= \frac{V_0}{27\alpha^2}\frac{1}{\sigma(\phi)^2} \left(1-\sqrt{1+6\alpha(\sigma(\phi) -1)}\right)^2 \nonumber \\
    &\mathrel{\phantom{=}} \times \left(1+2\sqrt{1+6\alpha(\sigma(\phi)-1)}\right),
\end{align}
where $\alpha$ is a dimensionless parameter defined as $\alpha = 2\gamma/3\beta^2$, and $\sigma(s)$ is defined in Eqn.~\eqref{sigma_phi} with $s=\phi/M^{*}_{\mathrm{Pl}}$. In Fig.~\ref{fig:potential_vs_alpha}, we plot the potential of the extended Starobinsky model, as function of inflaton value $\phi$, for different values of the dimensionless parameter $\alpha$. In the presence of the $R^3$ term, the potential plateau of this model only extends to the maximum at $\phi_{\rm max}$ and the universe slow-rolls to the lower field $\phi$. There are two possible ground states on both sides of the slow-roll plateau and we will consider only the slow-roll inflation to the low-field side. This potential reduces to the form of the Starobinsky model in the limit $\alpha \to 0$. Since $\gamma$, the $R^3$ coupling constant, is relatively small compared to $\beta$, the $R^2$ coupling constant, we can treat $\alpha$ in the potential as a small perturbation. Expanding the potential expression~\eqref{potential_exStar} in a series of $\alpha$, we obtain

\begin{align}
    V_{\rm exStar} &= V_0\bigg[\frac{(\sigma-1)^2}{\sigma^2}-\frac{(\sigma-1)^3\alpha}{\sigma^2} \nonumber \\ 
    &\mathrel{\phantom{=V_0\bigg[}} +\frac{9(\sigma-1)^4\alpha^2}{4\sigma^2}\nonumber \\
    &\mathrel{\phantom{=V_0\bigg[}}  + \frac{27(\sigma-1)^5\alpha^3}{4\sigma^2}+ \mathcal{O}(\alpha^4)\bigg].
\end{align}
This expansion is equivalently an infinite sum of terms proportional to $\alpha(\sigma-1)$. To ensure a valid approximation, we require $\sigma < 1+(1/\alpha)$. Additionally, from the potential in Eqn~\eqref{potential_exStar}, the field $\sigma$ must also satisfy $\sigma < 4+\sqrt{2/\alpha}$, which corresponds to its maximum where ${\rm d}V(\sigma)/{\rm d}\sigma=0$, ensuring that the inflation remains in the quasi-de Sitter phase. Since $\alpha \ll 1$, the value of $\sigma$ naturally falls within this bound, justifying approximation $\alpha(\sigma-1) \ll 1$. Henceforth, we will consider only the $\mathcal{O}(\alpha)$ terms in the potential~\eqref{potential_exStar}, and the potential can be approximated to the following expression 
\begin{equation}\label{approx_potential_exStar}
    V_{\rm exStar}(\phi) = V_{\rm Star}(\phi)\left[1-\alpha(\sigma(\phi)-1)\right],
\end{equation}
where $V_{\rm Star}(\phi)$ is the potential of the Starobinsky model~\eqref{potential_Star} (for $\gamma = 0$). To determine the value of $\beta$ for the extended Starobinsky model, we also impose COBE normalization in Eqn.~\eqref{COBE_normalization}, similar to what we have done for the Starobinsky model. We can thus obtain an expression for the $R^2$ coupling, determined in terms of $e$-folds at the horizon exit $N_e$ and the dimensionless parameter $\alpha$,
\begin{equation}
    \beta \simeq \frac{N_e^2}{(0.027)^4 {M^{*}_{\rm Pl}}^2}\left(\frac{1}{3} + \frac{32}{81}\alpha N_e^2\right). \label{beta-constraint-ext}
\end{equation}

Then we define dimensionless Cartesian coordinates for the effective phase space:
\begin{align}
     x &= \frac{1}{2\sqrt{3}}\left(1-\frac{1}{\sigma}\right)\sqrt{1-\alpha(\sigma-1)},\label{dimless coord.exStar x}, \\ 
    y &= \sqrt{\frac{\beta}{6}}\frac{\dot{\phi}}{M_{\rm Pl}^*}\label{dimless coord.exStar y}.
\end{align}
Hence, the polar coordinates $(z,\theta)$ can be written as
\begin{align}
    z &\equiv \sqrt{x^2+y^2} = \sqrt{\beta}H, \\
     \tan\theta &\equiv \frac{y}{x} = \frac{\sqrt{2\beta}\dot{\phi}\sigma}{M_{\rm Pl}^{*}(\sigma -1)\sqrt{1-\alpha(\sigma-1)}}.
\end{align}

Again, using Eqn.~\eqref{eqn:motion} and Eqn.~\eqref{eqn:traj}, we can plot trajectories in the $\phi-\dot{\phi}$ plane to explore the effective phase space. In $(x,y)$ coordinates, the velocity vector $\textbf{v}= \dot{\textbf{x}}/\sqrt{\beta}$ using Eqn.~\eqref{eqn:traj} is
\begin{widetext}
\begin{equation}\label{vec_Cart_ex_Star}
    \dot{\textbf{x}} = \frac{y}{\sqrt{3}}
    \left[\frac{1}{2}\frac{\alpha(1-\sigma)}{\sqrt{1-\alpha(\sigma-1)}} + \frac{\sqrt{1-\alpha(\sigma-1)}}{\sigma}\right]\Hat{\textbf{x}}
    -\left[3y\sqrt{x^2+y^2} + \frac{1}{6}\left(1-\frac{1}{\sigma}\right)\left( \frac{1-\alpha(\sigma-1)}{\sigma}-\frac{\alpha(\sigma-1)}{2}\right)\right]\Hat{\textbf{y}},
\end{equation}
\end{widetext}
where $\sigma$ can be expressed as a function of $x$ in Eqn.~\eqref{dimless coord.exStar x}. By asymptotic expansion, we obtain
\begin{equation}
    \sigma(x) = \frac{1}{1-2\sqrt{3}x} + \alpha\frac{6x^2}{\left(1-2\sqrt{3}x\right)^3}.
\end{equation}
This is a good approximation for exploring the effective phase space and evaluating the conserved measure from Liouville's theorem. The Starobinsky model can be expressed through the extended Starobinsky model by setting $\alpha = 0$, so that our velocity vector~\eqref{vec_Cart_ex_Star} clearly turns into the form of the velocity vector in the Starobinsky model in Eqn.~\eqref{vec_Cart_Star} and the function $\sigma(x)$ also reduces to the same function of the Starobinsky model, which is $\sigma(x) = 1/(1-2\sqrt{3}x)$. 

\begin{figure*}
\begin{centering}
    \includegraphics[width=0.31\linewidth]{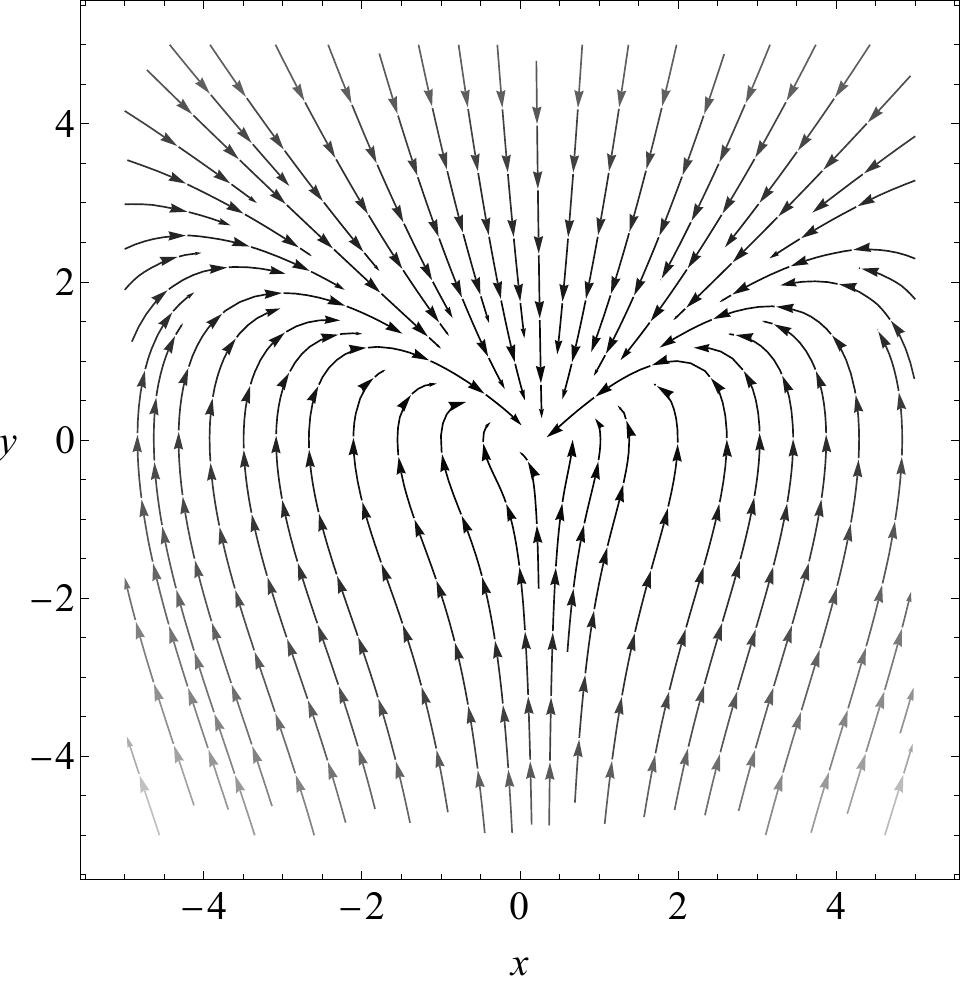}
    $\;\;\;\;$
    \includegraphics[width=0.31\linewidth]{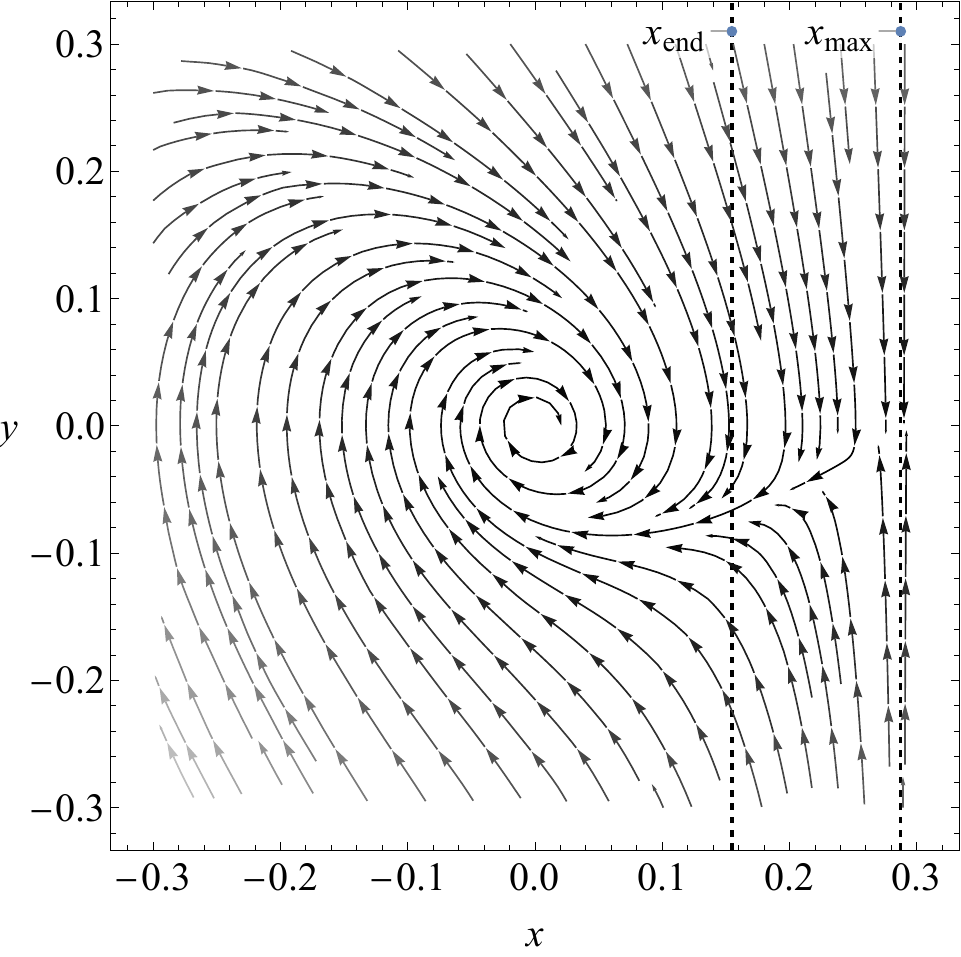}
    $\;\;\;\;$
    \includegraphics[width=0.31\linewidth]{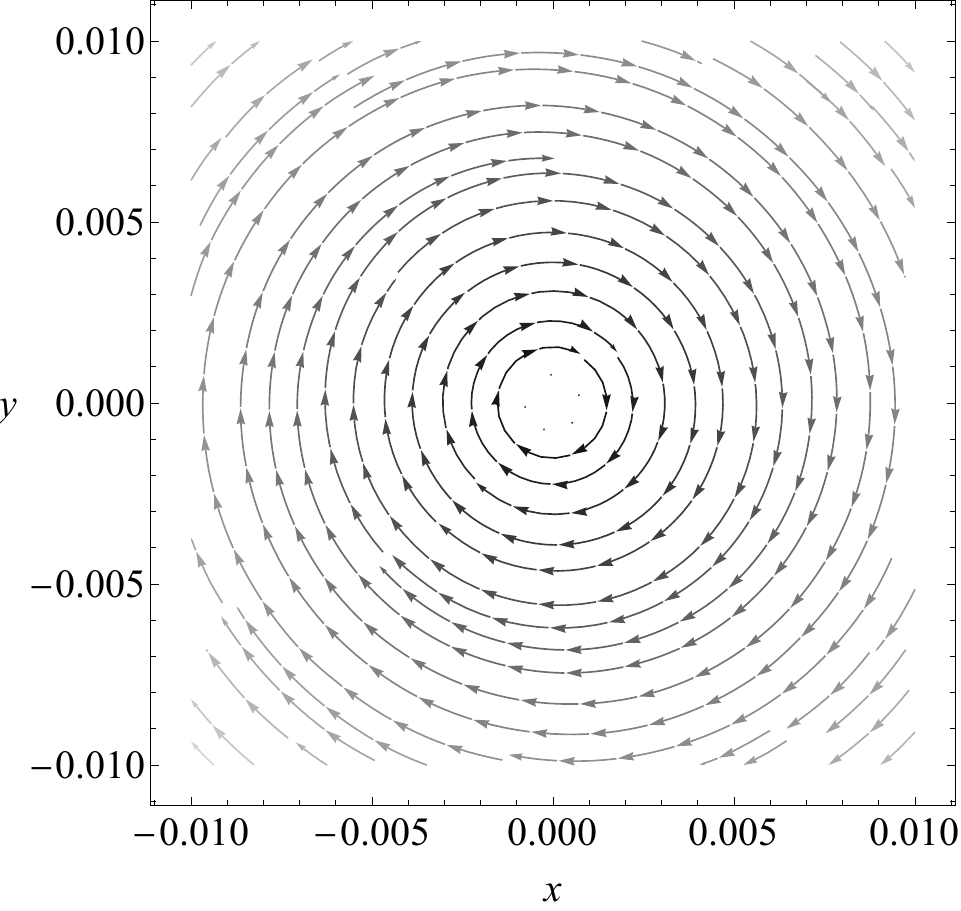}   
\end{centering}
    \caption{The stream plot of vector fields in the effective phase space illustrated by Eqn.~\eqref{vec_Cart_ex_Star} with $M^{*}_\mathrm{Pl} = 1$, $\beta = 0.1$, and $\alpha = 10^{-5}$. Similar to the Starobinsky model, there are three apparent attractors in the fixed angle large-field~(left), small-field~(right), and intermediate slow-roll region~(middle). Oscillatory attractor is apparent for small $x,y$ region in the right plot where reheating phase takes over. }
    \label{fig:vec_field_exStar}
\end{figure*}

To examine the effective phase space in the early universe, we simply transform the velocity vector~\eqref{vec_Cart_ex_Star} for Cartesian coordinate into polar coordinates. With series expansion with respect to $\alpha$ in the large $z$ limit, we obtain
\begin{widetext}
\begin{equation}\label{vec_polar_ex_Star}
    \dot{\mathbf{x}} = \left[-3z^2\sin^2\theta + \mathcal{O}(\alpha^2)\right] \Hat{\mathbf{z}} - \left[z^2\cos\theta(3\sin\theta-2)+\frac{z}{\sqrt{3}}+\frac{2z^2}{2\sqrt{3}z-\sec\theta}\alpha + \mathcal{O}(\alpha^2) \right]\Hat{\mathbf{\theta}}.
\end{equation}
\end{widetext}
This velocity vector also reduces to the form of Eqn.~\eqref{vec_polar_Star} for $\alpha \to 0$. The vector field in the effective phase space is plotted in Fig.~\ref{fig:vec_field_exStar}. Again, it has fixed-angle $\theta = \arcsin(2/3)$ apparent attractor for large field and the slow-roll attractor in the intermediate field region. For small field, there is oscillatory attractor representing reheating phase. 

\subsection{\texorpdfstring{Counting $e$-folding number in each trajectory}{Counting e-folds in each trajectory}}

Since we have seen in the Starobinsky model that the leading-order and next-to-leading-order calculations give almost identical results in the $e$-folds, we will consider only the leading-order slow roll in this extended Starobinsky model. The slow-roll parameters given by Eqn.~\eqref{slow-roll par} are
\begin{align}
    \epsilon_V &= \frac{((\sigma^2+\sigma-2)\alpha -2)^2}{3(\sigma-1)^2(1+\alpha(1-\sigma))^2}, \label{epsilon_V_exStar} \\
    \eta_V &= \frac{2\alpha\left(\sigma^3+3\sigma-4\right) +4(\sigma-2)}{3(\sigma-1)^2(\alpha(\sigma-1)-1)}.
\end{align}
Inflation ends when $\epsilon_V = 1$ and the number of $e$-folds stops counting at $\sigma_{\rm end} = 1+2/\sqrt{3}$ or $x_{\rm end} = (2-\sqrt{3})/\sqrt{3}$, for $\alpha = 0$. For extended Starobinsky model, we found that $\sigma_{\rm end}$ can be expressed in terms of $\alpha$ as
\begin{align}\label{sigma_end_exStar}
    \sigma_{\rm end} &= \frac{1}{4\alpha}\biggl[(3+\sqrt{3}+(7+3\sqrt{3})\alpha -\Big(12+6\sqrt{3}\nonumber \\
    &\mathrel{\phantom{=}} +(20+8\sqrt{3})\alpha+(36+18\sqrt{3})\alpha^2 \Big)^{1/2}\biggr].
\end{align}
Furthermore, we define $\sigma_{\rm max}$ as the value that gives the maximum value of $x$, denoted as $x_{\rm max}$. It can be determined by ${\rm d}x(\sigma)/{\rm d}\sigma = 0$ at $\sigma = \sigma_{\rm max}$, where $x(\sigma)$ is a function defined in Eqn~\eqref{dimless coord.exStar x}. We therefore obtain
\begin{equation}\label{sigma_max_exStar}
    \sigma_{\rm max} = \frac{1}{2} \left(\sqrt{\frac{8}{\alpha }+9}-1\right).
\end{equation}

The number of $e$-folds from the start at $\sigma$ to the end of inflation at $\sigma_{\rm end}$ can then be determined by Eqn.~\eqref{leading order N} combining with the expression~\eqref{sigma_phi}:
\begin{align}\label{eqNext}
    &N_1(\sigma) \nonumber \\
    &\;\;\;= \frac{3}{8}\left[\ln\left(\frac{(2-\alpha(\sigma^2+\sigma-2))^3}{\sigma^2}\right)\right. \\ \nonumber
    &\;\;\;+ \left.\frac{18\alpha+8}{\sqrt{9\alpha^2+8\alpha}}\tanh^{-1}\left(\sqrt{\frac{\alpha}{9\alpha+8}}(2\sigma + 1)\right)\right]\bigg|_{\sigma_{\rm max}}^{\sigma}.
\end{align}

\begin{figure}[t]
    \centering
    \includegraphics[width=1\linewidth]{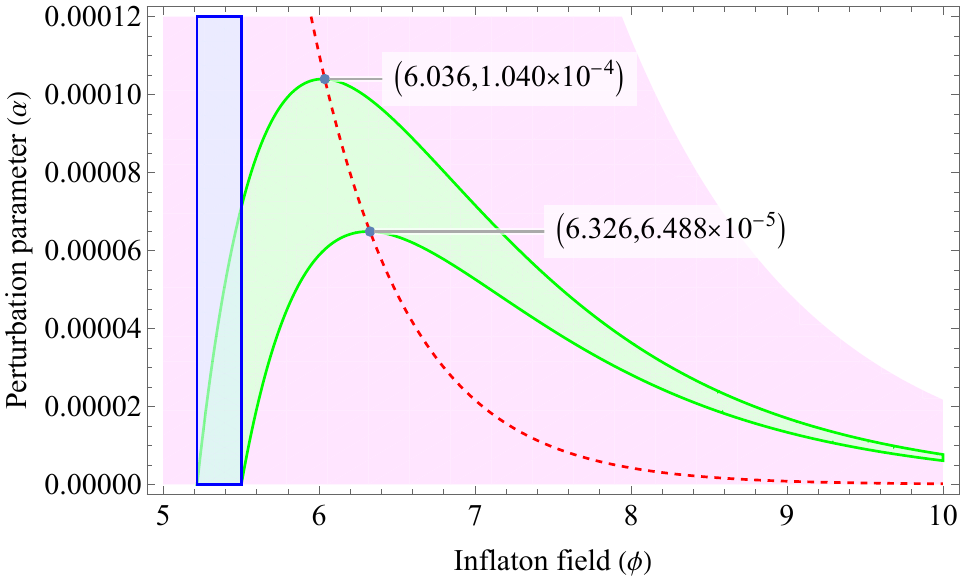}
    \caption{Constraints from the spectral index $n_{s}$ and the tensor-to-scalar ratio $r$ in the extended Starobinsky model as a function of $\phi<\phi_{\rm max}$. The green area indicates the allowed interval for $0.9067 < n_s < 0.9691$, the magenta area is the allowed region from the constraint on tensor-to-scalar ratio, $r < 0.036$, and the red dashed line represents the relation of $\phi_{\rm max}$ and $\alpha$. For comparison, the blue area corresponds to the observation constraints of the spectral index $n_s$ in the Starobinsky model. Constraint $r<0.036$ on the Starobinsky model is $\phi_{\rm UV}\gtrsim 2.4 M^{*}_{\rm Pl}$ and we do not express it here.}
    \label{fig: observational constraint exStar}
\end{figure}

Since the observables $n_s, r$ are related to the first-order slow-roll parameters in Eqs.~\eqref{n_s with slow-roll parameter} and~\eqref{r with slow-roll parameter}, the observational constraints given by Eqs.~\eqref{n_s constraint} and~\eqref{r constraint} can put constraints on the perturbation parameter $\alpha$ by using Eqn.~\eqref{sigma_max_exStar}. The allowed range of $\alpha$ from the observational constraints is shown in Figure~\ref{fig: observational constraint exStar}~(consistent with results from Ref.~\cite{Ivanov_2022}). Note that in the extended Starobinsky model, $\epsilon_{V}=0$ at $\phi_{\rm max}/M_{\rm Pl}^* = \sqrt{3/2}\ln\sigma_{\rm max}$ and thus $N(\phi_{\rm max})=\infty$, while in the Starobinsky model, $N(\phi\to\infty)=\infty$. The available range of $\alpha$ from Figure~\ref{fig: observational constraint exStar} is $1.04\times 10^{-4}\geq \alpha\geq 0$ for $\phi\leq \phi_{\rm max}$. If we require that the inflation starts at $\phi_{\rm max}$, the allowed range becomes $1.04\times 10^{-4}\geq \alpha\geq 6.488\times 10^{-5}$.

\subsection{\texorpdfstring{The expectation value of $e$-folding number and the probability distribution}{The expectation value of e-folds and the probability distribution}}
From Eqn.~\eqref{eqNext}, we obtain counting $e$-folding number for a given trajectory with initial angle $\theta$ on the Planck surface. As we have done in~\ref{sec:exp_e_folds_Star}, we calculate the expectation value of the number of $e$-folds from every possible trajectories by applying the conserved measure~\eqref{pdf}. First, we determine the conserved measure from Eqn.~\eqref{conservation}. In early universe, imposing $z \gg 1$ in Eqn.~\eqref{vec_polar_ex_Star} to obtain
\begin{equation}\label{vec_polar_early_ex_Star}
\begin{aligned} 
    \dot{\mathbf{x}} &\simeq -3z^2\sin^2\theta \Hat{\mathbf{x}} \\
    &\mathrel{\phantom{=}} -\left[z^2\cos\theta(3\sin\theta-2)+\frac{z}{\sqrt{3}}(1+\alpha)\right]\Hat{\mathbf{y}}.
\end{aligned}
\end{equation}

Interestingly, the contribution from $R^3$ term which is proportional to $\alpha$ only comes in at subleading order $\mathcal{O}(z)$. We thus need to keep subleading-order terms to examine the effects of the $R^{3}$ contribution. To solve this partial differential equation, we take the solution of the measure $\omega$ to the leading and subleading order as follow:
\begin{equation}\label{omega_exStar}
    \omega(z,\theta) = R_0(z)\Theta_0(\theta) + R_1(z)\Theta_1(\theta).
\end{equation}
The first term is the leading-order term and we expect its solution to be the same as Eqn.~\eqref{omega_Star}. It also implies that $R_0(z) \gg R_1(z)$. Substitute the ansatz into Eqn.~\eqref{conservation} with the velocity vector~\eqref{vec_polar_early_ex_Star}, we simply obtain the equation for large $z$ as
\begin{align}\label{pde_ex_Star}
    0 &= 3\sin^2\theta\partial_z\left[z^3\left(R_0\Theta_0 + R_1\Theta_1\right)\right] \nonumber\\
        &\mathrel{\phantom{=}} + \partial_\theta\left[z^2\cos\theta(3\sin\theta-2)\left(R_0\Theta_0 + R_1\Theta_1\right)\right] \nonumber\\
        &\mathrel{\phantom{=}} + \frac{z}{\sqrt{3}}(1+\alpha)R_0\partial_\theta \Theta_0,
\end{align}
where the smallest term $zR_1$ is negligible due to the large-$z$ limit. A solution for the measure has the requirement that it must be a periodic function in $\theta$. We obtain the solution to Eqn.~\eqref{pde_ex_Star} by considering the leading and subleading term separately.
For the leading term, 
\begin{align}
    0 &= 3\sin^2\theta\partial_z\left(z^3R_0\right)\Theta_0 \nonumber \\
        &\mathrel{\phantom{=}} + \partial_\theta\left(z^2\cos\theta(3\sin\theta-2)\Theta_0\right)R_0,
\end{align}
which is the same as Eqn.~\eqref{pde_ex_Star}. Therefore, the solutions can be written as in Eqn.~\eqref{radial_sol_Star} and~\eqref{aungular_sol_Star}.
For subleading term, we have
\begin{align}
    0 &= 3\sin^2\theta\partial_z\left(z^3R_1\right)\Theta_1 \nonumber\\
        &\mathrel{\phantom{=}} + \partial_\theta\left(z^2\cos\theta(3\sin\theta-2)\Theta_1\right)R_1 \nonumber\\
        &\mathrel{\phantom{=}} + \frac{z}{\sqrt{3}}(1+\alpha)R_0\partial_\theta \Theta_0.
\end{align}
To apply separation of variables to this equation, we assume that the solution of $R_1$ is $R_1 = R_0/z$. One can have
\begin{equation}
    3-m = \frac{\partial_\theta\left[ \cos\theta(3\sin\theta-2)\Theta_1 +\frac{1}{\sqrt{3}}(1+\alpha)\Theta_0\right]}{\Theta_1\sin^2\theta},
\end{equation}
which implies
\begin{widetext}
\begin{align}
    \Theta_1(\theta) &= \frac{1}{27}(1-\sin\theta)^{(m-4)/2}(1+\sin\theta)^{(m-8)/10}(2-3\sin\theta)^{-(3+4m)/15} \nonumber \\
    &\mathrel{\phantom{=}} \times \left[27C_1 + \frac{\sqrt{3}C(1+\alpha)(1+\sin\theta)^{3/10}\left((m-3)\cos2\theta+5m\sin\theta-3(8\sin\theta+m-6)\right)}{(2-3\sin\theta)^{9/5}\sqrt{1-\sin\theta}}\right].
\end{align}
\end{widetext}

The measure would again diverge at $\theta = \arcsin(2/3)$, where the fixed-angle apparent attractor is located in the large field. We then select $C_1 = 0$ and $m=5$ as our physical solution to obtain
\begin{equation}
    R_1(z) = \frac{C}{z^{7/3}},
\end{equation}
and
\begin{equation}\label{augular_sol_ex_Star}
    \Theta_1(\theta) = C\frac{(1+\alpha)(2\cos2\theta+\sin\theta+3)}{9\sqrt{3}(2-3\sin\theta)^{10/3}}.
\end{equation}
Demanding $\omega$ be positive everywhere, we take absolute value of $\omega$ in Eqn.~\eqref{omega_exStar} and obtain the probability distribution over the space of trajectories, parametrized by the angle $\theta$ on the Planck surface,
\begin{widetext}
\begin{equation}\label{approx_prob_dist_ex_Star}
    P(\theta)|_{H = M^{*}_{\rm Pl}} = C\left[
    \frac{(1-\sin\theta)^{2}\sin^2\theta}
    {\left|2-3\sin\theta\right|^{7/3}} + \frac{1}{{M^{*}_{\rm Pl}}\sqrt{\beta}}\frac{(1+\alpha)(2\cos2\theta+\sin\theta+3)\sin^2\theta}{9\sqrt{3}|2-3\sin\theta|^{10/3}}\right],
\end{equation}
\end{widetext}
where the normalization factor $C$ is evaluated by numerical integration over the physical region of the angle $\theta$.

Before computing expectation value of $e$-folds, we need to find a solution of $\sigma(x)$ from the definition of the dimensionless coordinates~\eqref{dimless coord.exStar x} to express the counting $e$-folds function $N(\sigma)$ in terms of $x$. It is simply a cubic equation that gives three roots, we choose the solution that presents the physical region where inflation occurs, i.e., region where inflaton slow-rolls to the lower field side towards oscillatory attractor~(reheating phase),
\begin{widetext}
\begin{equation}\label{sigma_x_exStar}
    \sigma(x) = \frac{1}{6\alpha}\left(2+6\alpha-24x^2-(1+i\sqrt{3})\xi - (1-i\sqrt{3})\frac{\left(1-12 x^2\right)^2-72\alpha x^2}{\xi}\right),
\end{equation}
where 
\begin{equation}
    \xi = \sqrt[3]{18\sqrt{\alpha^2x^2 \left(3x^2\left(9(3\alpha+4)\alpha-48 (\alpha +1)x^2+8\right)-1\right)}-18 x^2\left(9\alpha^2+6\alpha+96x^4-24(3\alpha+1)x^2+2\right)+1}.
\end{equation}
\end{widetext}

\begin{figure}[t]
    \centering
    \includegraphics[width=1\linewidth]{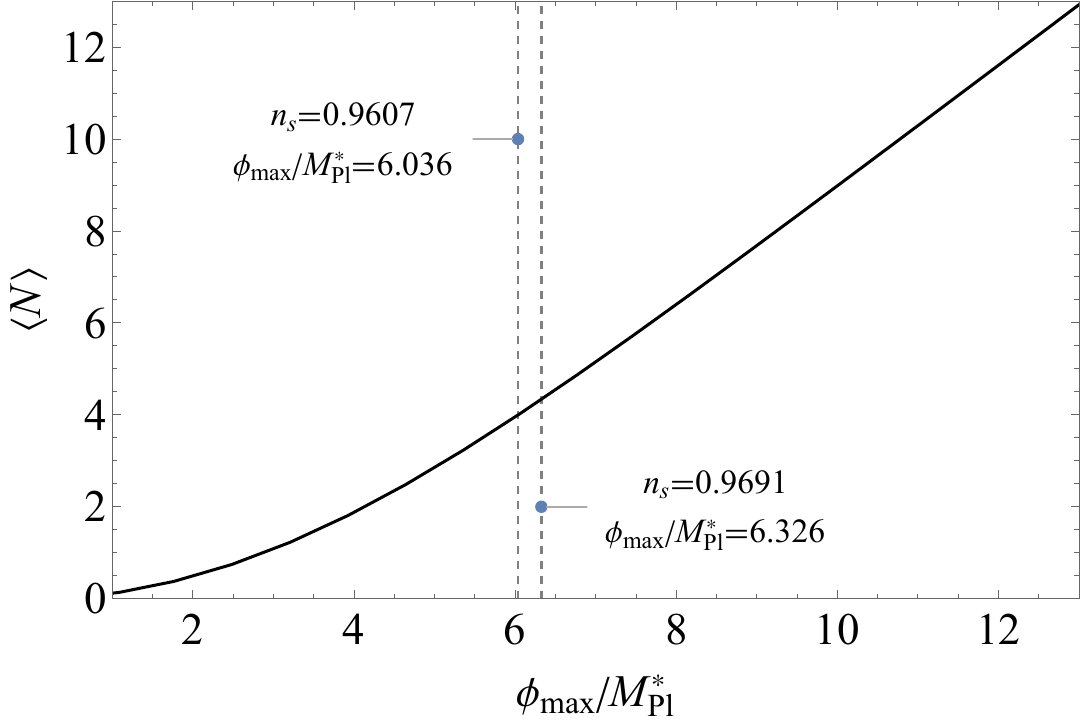}
    \caption{Expectation value of $e$-folds $\langle N \rangle$, numerically computed using the conserved measure~\eqref{approx_prob_dist_ex_Star} for the extended Starobinsky model versus the maximum inflaton $\phi_{\rm max}$.}
    \label{fig:e_folds_vs_phimax_exStar}
\end{figure}

Having the probability distribution~\eqref{approx_prob_dist_ex_Star}, we can now compute the expectation value of $e$-folds from the counting $e$-folding function $N_1(\theta)$ in the extended Starobinsky model by using Eqn.~\eqref{eqNext}, Eqn~\eqref{sigma_x_exStar}, and $x=M_{\rm Pl}^*\sqrt{\beta}\cos\theta$ on the Planck surface. We then evaluate the integral for the expectation value of $e$-folds over the interval $(\theta_{\rm max}, \theta_{\rm end}) \cup (2\pi - \theta_{\rm end}, 2\pi- \theta_{\rm max})$, where $\theta_{\rm max}$ and $\theta_{\rm end}$ are expressed in terms of $\sigma_{\rm max}$ and $\sigma_{\rm end}$, as defined by Eqs.~\eqref{sigma_max_exStar} and~\eqref{sigma_end_exStar}, respectively. The relationship between $\sigma$ and $\theta$ is given by
\begin{equation}\label{theta_sigma_relation}
    \theta = \arccos\left[\frac{\sqrt{1-\alpha(\sigma-1)}}{2\sqrt{3}M_{\rm Pl}^*\sqrt{\beta}}\left(1-\frac{1}{\sigma}\right)\right].
\end{equation}
Therefore, the expectation values of $e$-folds, $\langle N \rangle$, numerically computed using the conserved measure~\eqref{approx_prob_dist_ex_Star} for different values of $\alpha$, assuming inflation starts at $\phi_{\rm max}$, are shown in Table~\ref{tab:N_average with alpha_exStar}. $\langle N \rangle$ versus $\phi_{\rm max}$ are plotted in Fig.~\ref{fig:e_folds_vs_phimax_exStar}.

\begin{table}[t]
    \caption{The expectation value of $e$-folds achieved by all trajectories in the extended Starobinsky model over the allowed interval $(\phi_{\rm end},\phi_{\rm max})$, evaluated for different values of the parameter $\alpha$ satisfying the scalar spectral index $n_s$ constraint.}
    \centering
    $$\begin{array}{||c|c|c|c||}
    \hline
        \alpha & \phi_{\rm max}/M_{\rm Pl}^* & \phi_{\rm end}/M_{\rm Pl}^* & \langle N_1 \rangle \\[1ex]
    \hline \hline
        10^{-4} & 6.060 & 0.940 & 4.025  \\ \hline
        9.0\times10^{-5} & 6.125 & 0.940 & 4.101  \\ \hline
        8.0\times10^{-5} & 6.197 & 0.940 & 4.185  \\ \hline
        7.0\times10^{-5} & 6.279 & 0.940 & 4.282  \\ \hline
        6.5\times10^{-5} & 6.325 & 0.940 & 4.336  \\ \hline        
    \end{array}$$
    \label{tab:N_average with alpha_exStar}
\end{table}

\begin{table}[t]
    \caption{The probability of achieving $N_1>N_0$ for different values of the dimensionless parameter $\alpha$ with $N_0=50, 60$ in the region $(\phi_{N_1=50},\phi_{\rm max})$ and $(\phi_{N_1=60},\phi_{\rm max})$ respectively.}
    \centering
    $$\begin{array}{||c|c|c|c|c||}
    \hline
        \text{\multirow{2}{*}{$\alpha$}} & \multicolumn{2}{|c|}{\phi_{N_1 = N_0}/M_{\rm Pl}^*} & \multicolumn{2}{|c||}{\hspace{1em} {\rm Prob}(N_1>N_0) \hspace{1em}} \\ [0.5ex] \cline{2-5} 
        & N_0 = 50 & N_0 = 60 & \hspace{0.5em} N_0 = 50 \hspace{0.5em} & \hspace{0.5em} N_0 = 60 \hspace{0.5em} \\
    \hline \hline
        10^{-4}          & 5.145 & 5.317 & 0.009 & 0.006 \\ \hline
        9.0\times10^{-5} & 5.154 & 5.330 & 0.010 & 0.007 \\ \hline
        8.0\times10^{-5} & 5.164 & 5.342 & 0.011 & 0.007 \\ \hline
        7.0\times10^{-5} & 5.173 & 5.356 & 0.011 & 0.008 \\ \hline
        6.5\times10^{-5} & 5.178 & 5.362 & 0.012 & 0.008 \\ \hline 
    \end{array}$$
    \label{tab:probability with alpha_exStar}
\end{table}

\begin{figure}[t]
    \centering
    \includegraphics[width=1\linewidth]{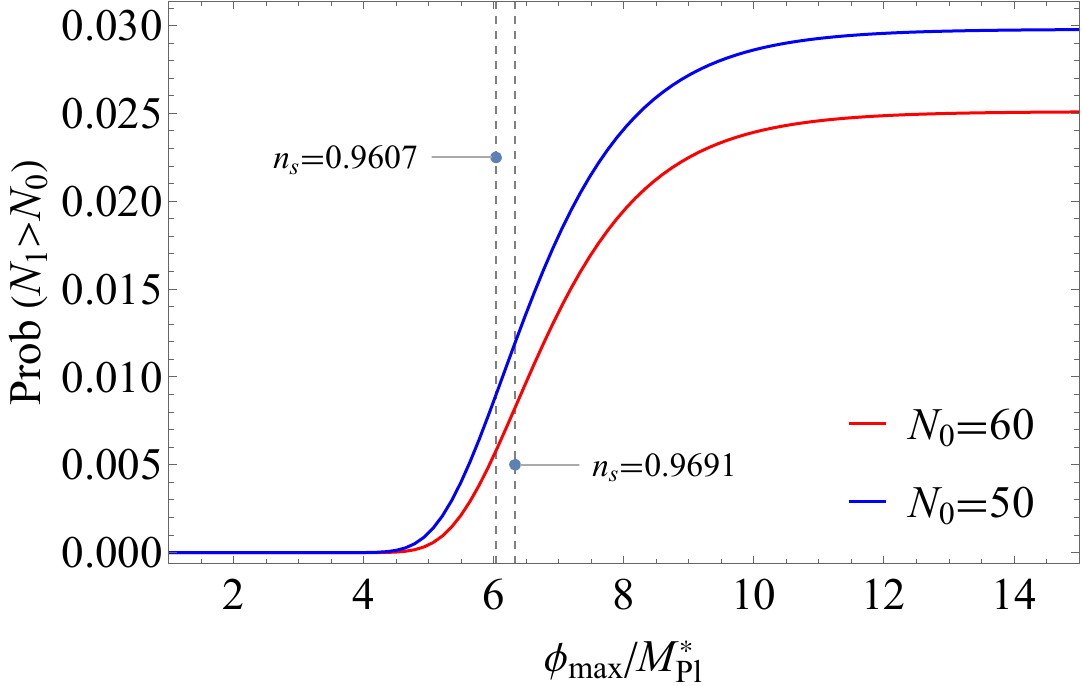}
    \caption{The probability of obtaining at least 50 and 60 $e$-folds as functions of the maximum inflaton $\phi_{\rm max}$ for the extended Starobinsky model.}
    \label{fig:fractionU_vs_phimax}
\end{figure}

Lastly, we evaluate how likely all possible initial conditions on the Planck surface with $e$-folding number given by Eqn.~\eqref{eqNext} would have $N > 50,60$, assuming inflation starts at $\phi_{\rm max}$. Namely, we compute the fraction of the trajectories within the subregion $(\theta_{\rm max}, \theta_{N_1=50,60})$, statistically weighed by the conserved measure~\eqref{approx_prob_dist_ex_Star}. The value of $\theta_{N_1=50,60}$ is determined numerically by solving $N_1(\sigma) = 50,60$, and then substituting the solution of $\sigma$ into Eqn.~\eqref{theta_sigma_relation}. The probability of achieving at least 50 and 60 $e$-folds during an inflationary period for different values of $\alpha$ is presented in Table~\ref{tab:probability with alpha_exStar}. Additionally, this probability is plotted in Fig.~\ref{fig:fractionU_vs_phimax} as a function of $\phi_{\rm max}$.  

\begin{figure}[t]
    \centering
    \includegraphics[width=1\linewidth]{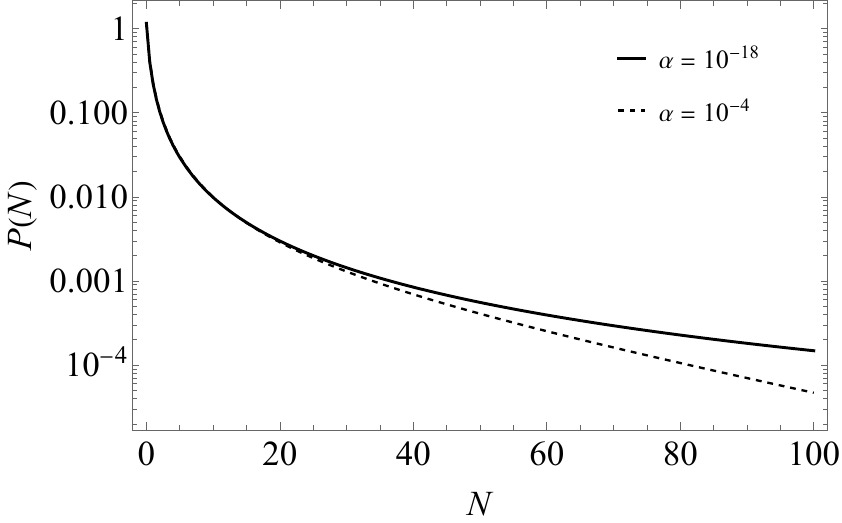}
    \caption{The probability distribution $P(N)$ of trajectories intersecting Planck surface versus $e$-folding number in the extended Starobinsky model for $\alpha = 10^{-18},10^{-4}$. $P(N)$ converges to the Starobinsky model as $\alpha\to 0.$}
    \label{fig:PN_vs_N_exStar}
\end{figure}

As the result, for $\alpha \to 0$, the probability $P(N>50)$ saturates at $0.0298$, and the probability $P(N>60)$ saturates at $0.0251$. Moreover, for a specific perturbation parameter $\alpha = 10^{-18}, 10^{-4}$, the probability distribution of trajectory space as a function of $e$-folds from $N_1$ is numerically plotted in Fig.~\ref{fig:PN_vs_N_exStar}. Similar to the Starobinsky model, the function decreases with increasing $N$ and approaches zero in the limit $N \to \infty$, resulting in the saturation of probability $P(N>N_0)$.

\section{Conclusions and Discussions}\label{Conclusion}

The Remmen-Carroll conserved phase space for flat universe has been constructed and explored in the Starobinsky and extended Starobinsky model of inflation. Under the slow-roll condition, the total $e$-folds along each trajectory is computed. The probability distribution of each inflationary trajectory intersecting the Planck surface is determined by the conserved measure. Among the trajectories which intersect the Planck surface, we find that the expectation value of $e$-folds $\langle N \rangle$ of the Starobinsky model depends crucially on the UV field $\phi_{\rm UV}$ where inflation starts, and it requires at least $\phi_{\rm UV}>50 M^{*}_{\rm Pl}$ for $\langle N \rangle >60$. Also $P(N>60)>0$ requires $\phi_{\rm UV}>5.5 M^{*}_{\rm Pl}$ in the Starobinsky model. The probability distribution $P(N)$ peaks at $N=0$ resulting in the saturation of $P(N>60)=0.0251$ for arbitrarily large $\phi_{\rm UV}$. Note that even at super-Planckian field $\phi_{\rm UV}=50 M^{*}_{\rm Pl}$, the energy density from the potential is still sub-Planckian,

\begin{equation}
    V(\phi_{\rm UV}=50 M^{*}_{\rm Pl}) = \frac{ M^{*2}_{\rm Pl}}{4\beta} = 1.1\times 10^{-10} M^{*4}_{\rm Pl},
\end{equation}
where we have used Eqn.~\eqref{potential_Star} and \eqref{beta_constraint}. The energy density actually saturates to this sub-Planckian value for arbitrarily large field. Also, the Hubble parameter is always considerably sub-Planckian as shown in Fig.~\ref{fig:Hubble_SR_vs_phi}. The inflation is still sub-Planckian and semi-classical even at the super-Planckian cutoff $\phi_{\rm UV}=50 M^{*}_{\rm Pl}$ and beyond. This is guaranteed by the slow-roll condition which suppresses the kinetic energy of the field via small $\dot{\phi}$. Quantum gravity graviton-exchange scattering occurs at large Planckian momentum which requires large $\dot{\phi}$, and not the field value itself. 

The only constraint on the field value comes from constraint on $N(\phi)$ via the $n_{s}$ constraint. Under such constraint, $\phi_{\rm UV}\in [5.22,5.50]M^{*}_{\rm Pl}$, the expectation value becomes $\langle N \rangle \simeq 3.5 - 4$ for trajectories intersecting the Planck surface. Namely, only a small fraction of the trajectories crossing the Planck surface reaches $N\geq 60$.  

For extended Starobinsky model with additional $R^3$ term parametrized by the coupling parameter $\alpha$, the potential plateau where slow-roll occurs is truncated to $\phi \leq \phi_{\rm max}$ as shown in Fig.~\ref{fig:potential_vs_alpha}, resulting in a limited period of inflation. For example, for $\alpha = 10^{-4}$, the expectation value drops to $\langle N \rangle = 4.03$ while for $\alpha = 10^{-35}$, $\langle N \rangle = 61.4$~(where $\phi_{\rm UV}\simeq 50 M^{*}_{\rm Pl}$) converging to the corresponding value in the Starobinsky model. The probability $P(N>60)$ for $\alpha > 0$ is also less than the saturation value $P(N>60)({\rm Starosbinky})=0.0251$ in the Starobinsky model. Similar to the Starobinsky model, Remmen-Carroll phase space reveals that only a small fractions of trajectories intersecting Planck surface could reach $N\geq 60$. 

As observed in Ref.~\cite{Remmen_2014} for quadratic and cosine inflaton potentials, there are two apparent attractors in the Remmen-Carroll phase space, one is the slow-roll attractor at large field and another is the small-field reheating attractor. In both Starosbinsky and extended Starobinsky models, there are three apparent attractors found in the stream plot of velocity fields. For large field region, there is a fixed angle attractor at $\theta =\arcsin(2/3)$. For intermediate field, there is slow-roll attractor connecting between ultraviolet field region and the small-field region. In the small field region, there is apparent oscillatory attractor where reheating phase should begin as shown in Fig.~\ref{fig:vec_field_Star} and Fig.~\ref{fig:vec_field_exStar}. 

As shown in these phase spaces~(middle/right plots of Figure~\ref{fig:vec_field_Star} and \ref{fig:vec_field_exStar}), any trajectories starting at the Planck surface will converge to the slow-roll attractor, then gradually spiral down to the reheating attractor regardless of the initial conditions within the possible range~($[x_{\rm 0},x_{\rm end}]$ and $[x_{\rm max},x_{\rm end}]$ for Starobinsky and extended Starobinsky model respectively). Slow-roll inflation is thus ``natural'' within this phase space. However, among all trajectories only a fraction reaches sufficiently large e-folds, $N>50-60$ as quantified by $P(N)$ in Fig.~\ref{fig:fractionU_vs_cutoff},\ref{fig:P(N)_vs_N_Star},\ref{fig:fractionU_vs_phimax} and \ref{fig:PN_vs_N_exStar}. The probability $P(N)$ of the Starobinsky model depends crucially on the cutoff $\phi_{\rm UV}$ and it requires at least $\phi_{\rm UV}/M^{*}_{\rm Pl}>5.43-5.45$ for the e-folds to reach $N_{2,1}\geq 60$. This is still consistent within the constraints from $n_{s}$ which requires $5.22<\phi_{\rm UV}/M^{*}_{\rm Pl}<5.50$. 

Given a smooth homogeneous and isotropic initial conditions with low entropy, even though it is natural to have a slow-roll inflation, it is unlikely that the Starobinsky inflation would reach 60 e-folds~(the probability is only 0.5\%, see Figure~\ref{fig:fractionU_vs_cutoff}) and completely resolve the horizon problem. The extended Starobinsky model has similar characteristic, only approximately $0.5\%$~(Figure~\ref{fig:fractionU_vs_phimax}) of trajectories reaches 60 e-folds for the values of inflaton cutoff satisfying observational constraints $6.036<\phi_{\rm max}/M^{*}_{\rm Pl}<6.326$ where $4.02<\langle N \rangle <4.34 $. For arbitrarily high inflaton cutoff, both models give $P(N>60)=0.025$, or only 2.5\% of the trajectories reach more than 60 e-folds.\\

\begin{acknowledgments}
TR and PB are supported in part by National Research Council of Thailand~(NRCT) and Chulalongkorn University under Grant N42A660500. 
\end{acknowledgments}

\appendix

\bibliography{efolds_inflation}
\bibliographystyle{apsrev4-1} 

\end{document}